\documentclass[aps, prl, 10pt, twocolumn, superscriptaddress,noshowpacs, preprintnumbers, longbibliography, bibnotes,hyperref,floatfix]{revtex4-1}

\usepackage[colorlinks=true,breaklinks=true]{hyperref}
\hypersetup{allcolors=[rgb]{0.0 0.0 1.0},linkcolor=[rgb]{0.75 0.05 0.05}}
\usepackage[dvipsnames]{xcolor}
\usepackage{mathtools}
\usepackage{amsmath}
\usepackage{amsfonts}
\usepackage{amssymb}
\usepackage{bbold}
\usepackage{multirow}
\usepackage{bm}
\usepackage{hyperref}
\long\def\exclude#1{}
\usepackage{orcidlink}
\usepackage{graphicx}    
\usepackage[caption=false]{subfig}
\usepackage{booktabs}    
\usepackage{adjustbox}   
\usepackage{soul}

\begin{document}

\title{
NuSTAR bounds on radiatively decaying particles from M82 
}

\author{Francisco R. Cand\'{o}n~\orcidlink{0009-0002-3199-9278}}
\affiliation{Departamento de F\'isica Te\'orica,
			Universidad de Zaragoza,
			C. Pedro Cerbuna 12,
			50009 Zaragoza,
			Spain}
\affiliation{Centro de Astropart\'iculas y F\'isica de Altas Energ\'ias (CAPA), Universidad de Zaragoza, 50009 Zaragoza, Spain}

\author{Damiano F.\ G.\ Fiorillo \orcidlink{0000-0003-4927-9850}}
\affiliation{Deutsches Elektronen-Synchrotron DESY,
Platanenallee 6, 15738 Zeuthen, Germany}

\author{Giuseppe Lucente
\orcidlink{0000-0003-1530-4851}}
\affiliation{SLAC National Accelerator Laboratory, 2575 Sand Hill Rd, Menlo Park, CA 94025}

\author{\\Edoardo Vitagliano
\orcidlink{0000-0001-7847-1281}}
\affiliation{Dipartimento di Fisica e Astronomia, Università degli Studi di Padova,
Via Marzolo 8, 35131 Padova, Italy}
\affiliation{Istituto Nazionale di Fisica Nucleare (INFN), Sezione di Padova,
Via Marzolo 8, 35131 Padova, Italy}

\author{Julia K. Vogel
\orcidlink{0000-0002-5850-5517}}
\affiliation{Fakult\"{a}t f\"{u}r Physik, TU Dortmund, Otto-Hahn-Str. 4, Dortmund D-44221, Germany}

\begin{abstract}
Axions and other putative feebly interacting particles (FIPs) with a mass of tens to several hundreds of keVs can be produced in stellar cores with a Lorentz boost factor $E_a/m_a\lesssim 10$. Thus, starburst galaxies such as M82 are efficient factories of slow axions. Their decay $a\rightarrow\gamma\gamma$ would produce a large flux of X-ray photons, peaking around $100$~keV and spread around the galaxy by an angle that can be relatively large. We use observations of the Nuclear Spectroscopic Telescope Array (NuSTAR) mission to show that the absence of these features can constrain $30-500$~keV axion masses into uncharted regions for axion-photon coupling of $g_{a\gamma}\sim 10^{-10}-10^{-12}\,\rm GeV^{-1}$. Our argument can be applied to other heavy FIPs and astrophysical sources that are hot enough to produce them, yet cold enough to avoid large boost factors which slow down the decay.
\end{abstract}

\date{\today}

\maketitle

\textbf{\textit{Introduction.}}---Axions, sterile neutrinos, and other hypothetical feebly interacting particles (FIPs) with a mass in the keV-GeV range have increasingly been in the spotlight in recent years~\cite{Chang:2016ntp,Hardy:2016kme,
Jaeckel:2017tud,Chang:2018rso,
DeRocco:2019njg,Croon:2020lrf,
Hoof:2022xbe,
Camalich:2020wac,Caputo:2021kcv,
Caputo:2021rux,Caputo:2022mah,Ferreira:2022xlw,Caputo:2022rca,Fiorillo:2022cdq,Carenza:2023old,
Manzari:2023gkt,Lella:2023bfb,Diamond:2023scc,
Akita:2023iwq,Muller:2023pip,
Fiorillo:2023cas,Fiorillo:2023ytr,Fiorillo:2024upk,Telalovic:2024cot,Hardy:2024gwy}. While the most stringent bounds on lighter FIPs often rest on stellar cooling arguments (see e.g.~\cite{Raffelt:1996wa,Caputo:2024oqc,Carenza:2024ehj}), as they would affect the standard evolution of stars, the existence of heavier FIPs is constrained by the null observation of the daughter particles produced by their decay. 
One representative case of such FIPs is an axion with a two-photon coupling $\mathcal{L}_{a\gamma\gamma}=g_{a\gamma}a\, \mathbf{E}\cdot \mathbf{B}$. Here we use natural units $\hslash=c=1$, so that $g_{a\gamma}$ has dimension (energy)$^{-1}$. Barring cosmological bounds (which depend on different parameters, such as the reheating temperature value~\cite{Depta:2020zbh,Balazs:2022tjl,Langhoff:2022bij}), strong constraints for axion masses $m_a$ between the eV and the MeV scale come from the observed cooling of horizontal-branch stars~\cite{Ayala:2014pea,Dolan:2022kul}, from the possible decay of axions produced by main sequence stars~\cite{Nguyen:2023czp}, and of axions
gravitationally trapped around the Sun (``solar basin'')~\cite{VanTilburg:2020jvl,Beaufort:2023zuj}.

At heavier masses, above the MeV range, the best probe is offered by the remnant of core-collapse supernovae (SNe), an extremely hot ($T\simeq 30\,\rm MeV$) and dense ($\rho\simeq 10^{14}\,\rm g/cm^3$) proto-neutron star~\cite{Raffelt:1996wa}. Axions as heavy as $\mathcal{O}(1\,\rm GeV)$ could have been copiously produced in SN~1987A through Primakoff effect and coalescence. The photons produced by their subsequent decay $a\rightarrow \gamma\gamma$ would have shown up in X-ray and gamma-ray observations of SN~1987A realized with the Solar Maximum Mission~\cite{Jaeckel:2017tud,Caputo:2021rux,Hoof:2022xbe} and the Pioneer Venus Observatory~\cite{Diamond:2023scc}. Other constraints from astrophysical transients also get weaker for smaller masses, e.g. the bounds from low-energy SNe~\cite{Caputo:2022mah}, and from the electromagnetic signal of the GW~170817 event~\cite{Diamond:2023cto,Dev:2023hax}. The general lesson is that the most constraining bounds are obtained for axions light enough to be produced, and heavy enough that the Lorentz boost does not impede their decay.

\begin{figure}
    \centering
\includegraphics[width=0.48\textwidth]{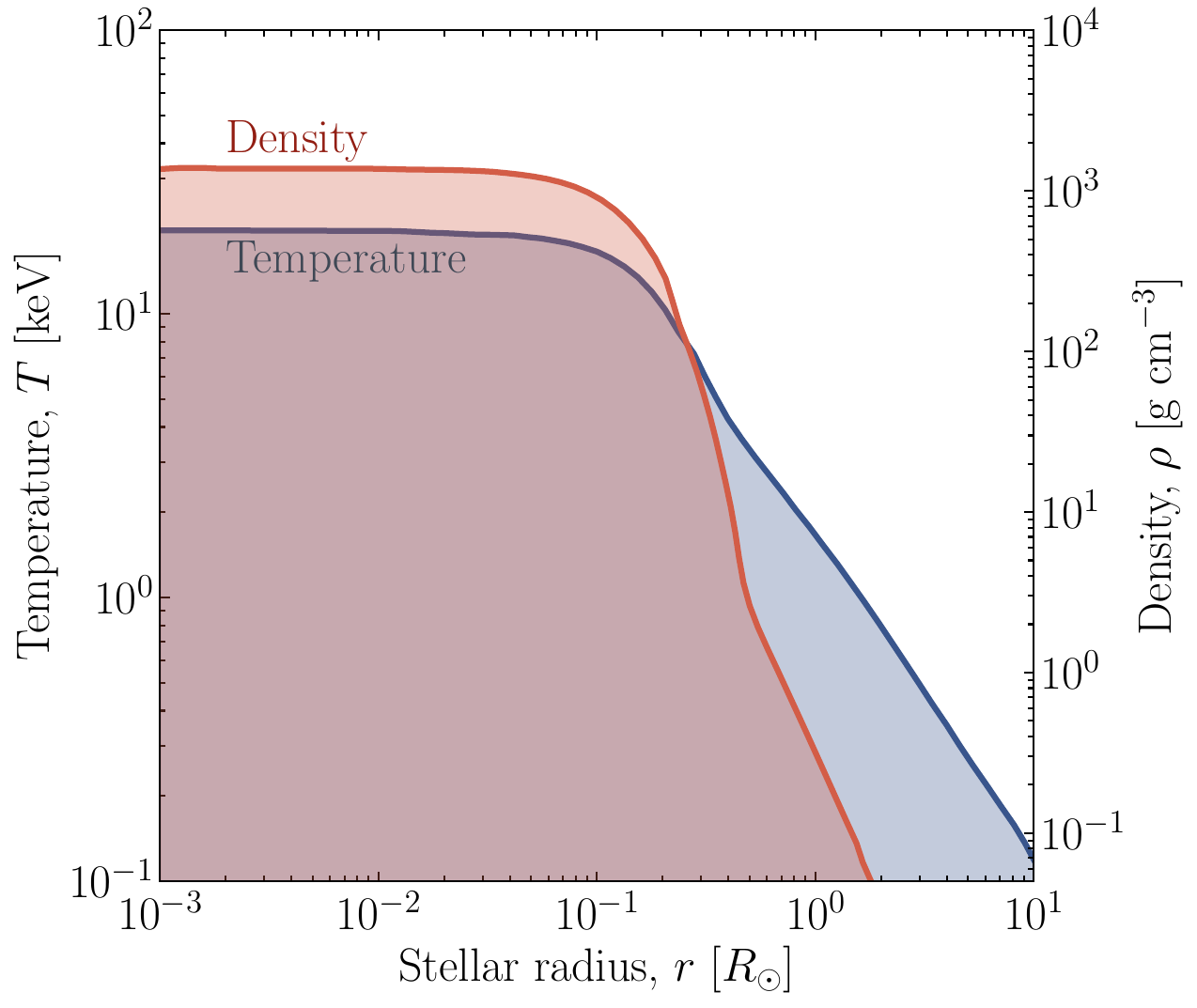} 
\caption{Temperature and density profile for a star with an initial mass of $20~M_\odot$ at an age $\sim 8.7$ Myr, as a function of stellar radius in units of the solar radius $R_\odot$. Such stars dominate the emission of axions from M82, since lighter stars attain smaller temperatures, while heavier stars are much less numerous.}
\label{fig:fig1}
\end{figure}

\begin{figure*}[t!]
    \includegraphics[width=\textwidth]{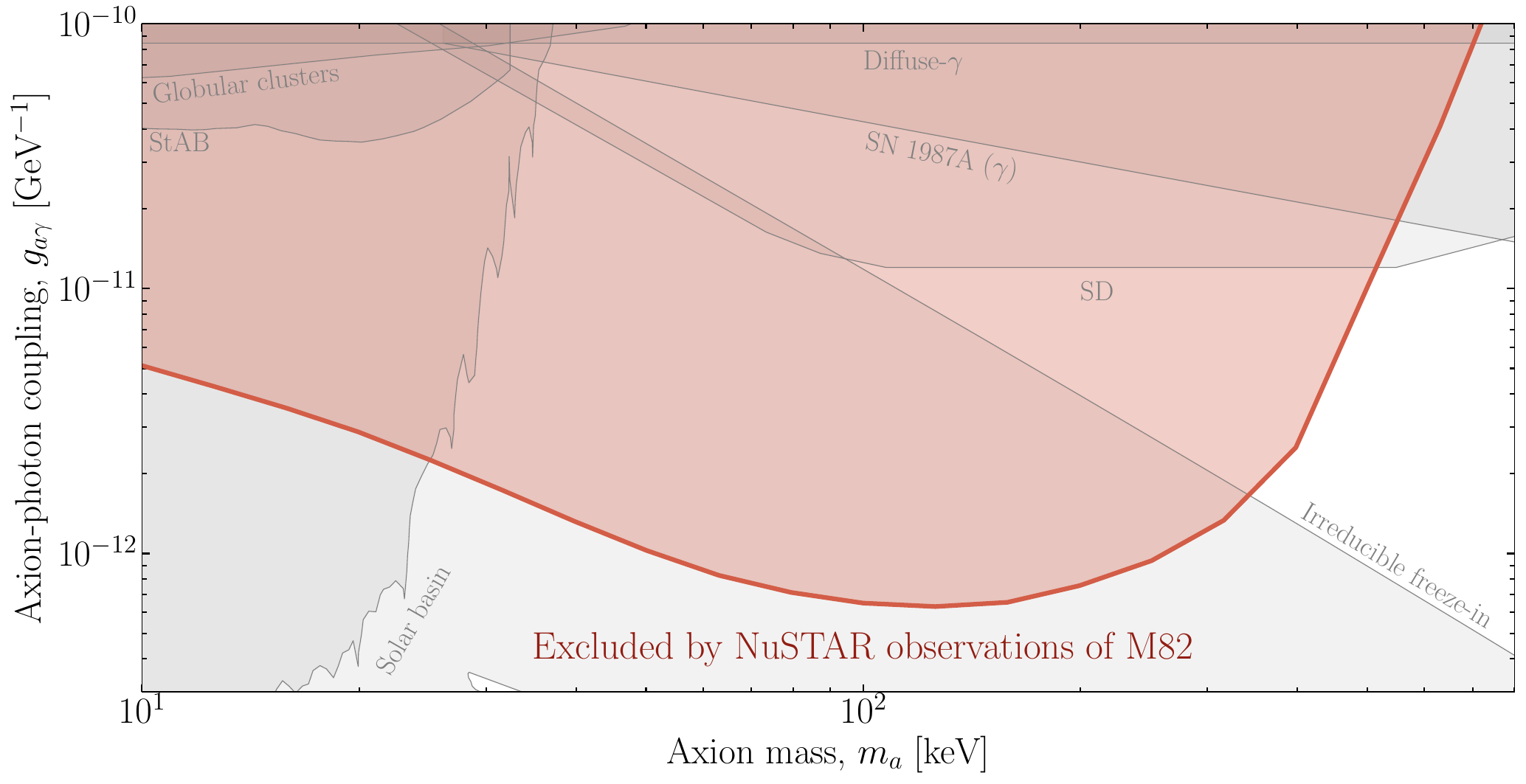}
    \caption{Novel bounds on axions produced in M82 and decaying to photons. We show previous bounds from the literature~\cite{Dolan:2022kul,Nguyen:2023czp,Beaufort:2023zuj,Hoof:2022xbe,Langhoff:2022bij,Caputo:2021rux}, all extracted from Ref.~\cite{AxionLimits}, in gray. We also show the bounds from spectral distortions (SDs) taken from Ref.~\cite{Langhoff:2022bij}; see main text for a discussion of these bounds.}\label{fig:bounds}
\end{figure*}

Here, we point out that this lesson can be extended to lower masses $m_a\lesssim 1\rm\, MeV$, by looking at sources hot enough to produce the axion, but with temperatures much smaller than the remnant of core-collapse SNe, so their Lorentz boost is smaller. Starburst galaxies (SBGs) meet exactly this requirement, due to their intense star-forming activity and their large number of stars reaching temperatures up to several tens of keV. This is quite visible in Fig.~\ref{fig:fig1}, showing the temperature and density profile for a typical star in the mass range that dominates the aggregated emission from the SBG.

Perhaps the most well-known SBG is M82, also known as the Cigar Galaxy. We show here that the axions produced in the stellar cores within the SBG should decay on their way to Earth and produce a photon flux peaking around a hundred keV, and with a characteristic angular distribution that can be either narrowly centered around M82, or widened up to even several arcminutes around the source due to the delayed decay of the mildly relativistic axions. Therefore, we can use the observations of M82 from the Nuclear Spectroscopic Telescope Array (NuSTAR) telescope~\cite{NuSTAR:2013yza}, primarily in the energy range of $E_\gamma$ between $30-70$~keV, to constrain the magnitude of this flux. By this strategy, we obtain the bounds shown in Fig.~\ref{fig:bounds}, which exclude a new window of the parameter space as large as one order of magnitude in coupling.

\textbf{\textit{Stellar population and axion production.}}---Axions interacting with photons can be produced in stars via two processes. For $m_a \lesssim T$, the only relevant mechanism is the Primakoff effect $\gamma + Ze \to Ze + a$~\cite{Dicus:1979ch,Raffelt:1985nk,DiLella:2000dn,Carenza:2020zil,Lucente:2020whw,Caputo:2022mah}, in which thermal photons are converted into axions in the electrostatic field of charged particles. For larger masses, the photon coalescence $\gamma \gamma \to a$~\cite{Carenza:2020zil,Lucente:2020whw,Caputo:2022mah} becomes important.

To compute the axion emission spectra from M82, we evaluate the axion production from individual stars, and then determine the aggregated signal from the stellar population of the Galaxy through the Star Formation History (SFH) and the Initial Mass Function (IMF), describing the stellar distribution according to their age and mass, respectively. As further discussed in Supplemental Material (SupM)~\cite{supplementalmaterial}, we compute the axion spectra from individual stars using radial profiles computed through the one-dimensional stellar evolution code Modules for Experiments in Stellar Astrophysics (MESA)~\cite{Paxton:2010ji,Paxton:2013pj} (release {\tt r23.05.1}). The dominant contribution to axion emission in M82 comes from high-mass stars (with mass $M\gtrsim 20~M_\odot$)~\cite{Ning:2024eky}, so we employ a grid of models between $20~M_\odot$ and $80~M_\odot$ spaced by $10~M_\odot$, evolved using the default MESA suite and in-list for high-mass stars, {\tt 20M\_pre\_ms\_to\_core\_collapse}, from pre-main-sequence to the onset of core collapse, providing us with stellar profiles at different ages, depending on the initial stellar mass. For instance, we show in Fig.~\ref{fig:fig1} the temperature (blue) and the density (red) as a function of the stellar radius for an initial $20~M_\odot$ star at an age $\sim 8.7$ Myr. We describe the stellar population in M82 using the SFH and the IMF adopted in Ref.~\cite{Ning:2024eky}. The SFH is a ``two-burst'' model described by $R_0\,e^{(t_{\rm burst}-t)/t_{\rm sc}}$ for $t < t_{\rm burst}$, with the normalization constant $R_0$ different for the two bursts and $t_{\rm sc}=1.0$~Myr the characteristic decay timescale~\cite{ForsterSchreiber:2003ft}.  The old burst is characterized by $t_{\rm burst}=9.0$~Myr and $R_0 = 31$~$M_\odot/{\rm yr}$, while the younger one by $t_{\rm burst}=4.1$~Myr and $R_0 = 18$~$M_\odot/{\rm yr}$~\cite{ForsterSchreiber:2003ft}. On the other hand, we assume the IMF for M82 to be $\propto M^{-2.35}$ at high mass and flat below $\sim 3M_\odot$, with a cutoff at 100~$M_\odot$~\cite{ForsterSchreiber:2003ft}. Thus, the total axion production spectrum is
\begin{equation}
  \frac{d\dot{N}_a}{dE_a} =   N_{\rm tot} \int dM_s\, dt_s\, {\rm IMF}(M_s)\,{\rm SFH}(t_s) \,  \frac{d\dot{N}^s_{a}}{dE_a} 
\end{equation}
where $N_{\rm tot}=1.8\times 10^{10}$~\cite{Ning:2024eky} is the total number of stars in M82 estimated using luminosity observations~\cite{NED} and color-mass-to-light ratio relations for disk galaxies from~\cite{1407.1839}, and $d\dot{N}^s_{a}/dE_a$ is the production spectrum from the single star with age $t_s$ and initial mass $M_s$. The determination of the axion production from each star is based on standard emission rates, as detailed in SupM~\cite{supplementalmaterial}. For the couplings we are interested in, the axion feedback on the stellar evolution is negligible since the axion luminosity is well below the stellar one. After integrating over the stellar population, we find that the dominant contribution to the axion production in M82 comes from stars with initial mass $20~M_\odot$, and that the photon coalescence is the main production process for $m_a \gtrsim 30$~keV.

\begin{figure}
    \centering
    \includegraphics[width=0.45\textwidth]{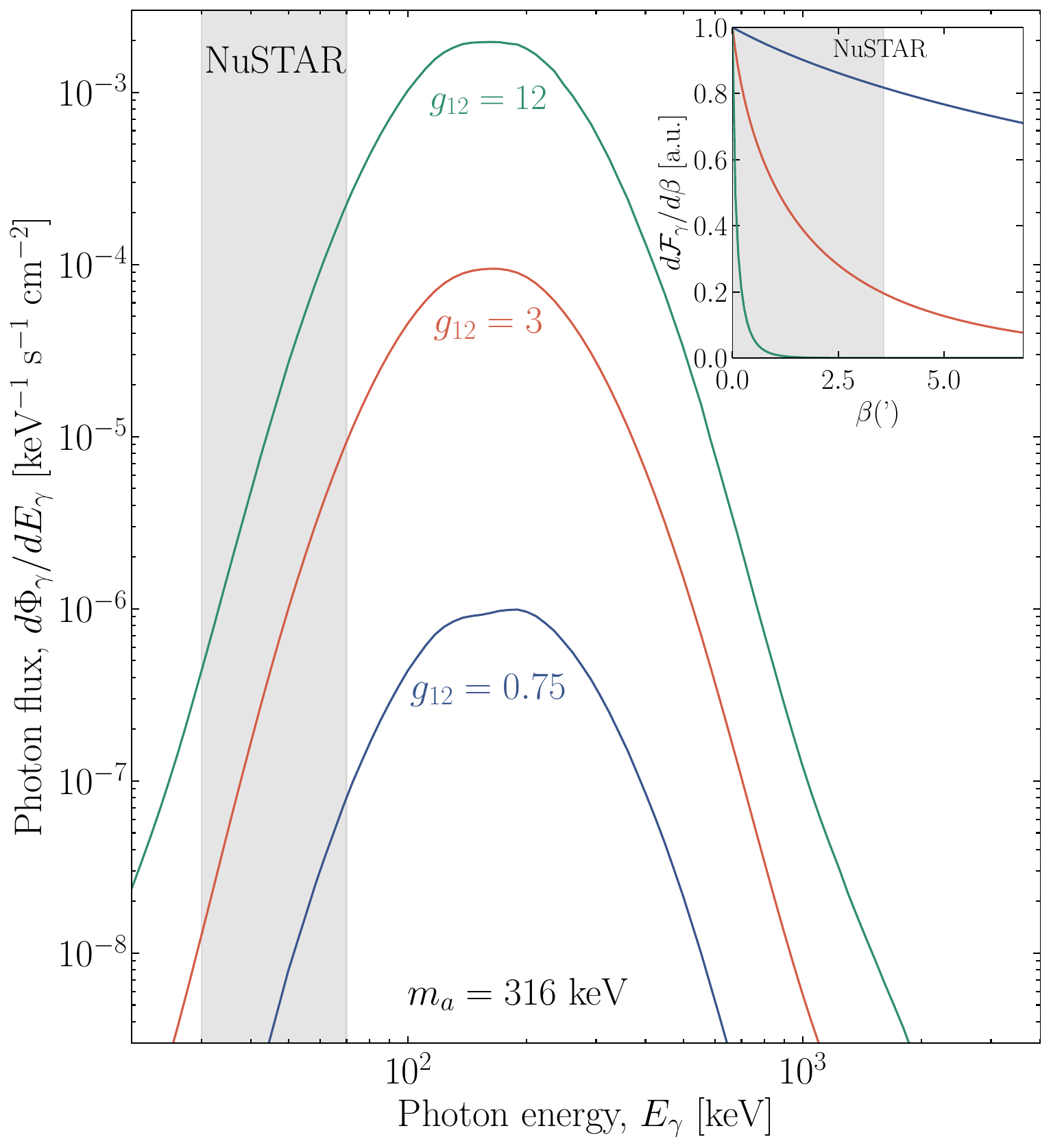}
    \caption{Energy (main plot) and angular (inset) distribution of the photons from axion decay. The angular and energy range used in the analysis of the NuSTAR observations is highlighted in gray. The energy distribution $d\Phi_\gamma/dE_\gamma$ is integrated over photons within the angular range used in the analysis; similarly, the angular distribution of the integrated energy flux $d\mathcal{F}_\gamma/d\beta$ is integrated over photons between $30$~keV and $70$~keV.}
    \label{fig:spectra}
\end{figure}

\textbf{\textit{Photon spectrum.}}---The decay of axions leads to the production of photons of comparable energies. When the decay is sufficiently slow, the decay length can become comparable with the distance between M82 and the Earth; in this case, photons can reach Earth even from directions away from M82. In the SupM~\cite{supplementalmaterial}, we deduce the differential flux $d\Phi_\gamma/dE_\gamma d\beta$ of these photons in energy $E_\gamma$ and angle from the source $\beta$ to be
\begin{align}
    \label{eq:photon_flux}
    \frac{d\Phi_\gamma}{dE_\gamma d\beta}&=\int_{E_{a,\mathrm{min}}}^{+\infty}\frac{2dE_a \cos\beta}{p_a}\frac{d\dot{N}_a/dE_a}{4\pi r}\\ \nonumber 
    &\Theta(\cos\beta-\cos\theta) \exp\left[-\frac{\Gamma_a m_a r \sin\beta}{p_a\sin\theta}\right]\frac{\Gamma_a m_a}{p_a\sin\theta}\,.
\end{align}
Here, $\Gamma_a=g_{a\gamma}^2 m_a^3/64\pi$ is the rest-frame decay rate of an axion with mass $m_a$, $E_a$ and $p_a$ are the axion energy and momentum, so that the velocity is $v_a=p_a/E_a$, $\Theta$ denotes the Heaviside theta, $r=3.8$~Mpc is the distance between M82 and the Earth and $\theta$ is the angle between the axion and photon direction at the moment of decay, with
\begin{equation}
    \cos\theta=\frac{1}{v_a}\left[1-\frac{E_a}{2E_\gamma}(1-v_a^2)\right].
\end{equation}
Equation~\eqref{eq:photon_flux} already includes a projection factor $\cos\beta$ for the photons impacting on a unit area of the detector, although for the small angles considered here it is irrelevant. We can clearly identify two extreme regimes: if $\Gamma m_a r/p_a\sin\theta\gg 1$, the flux is strongly suppressed at large angles. The axions are decaying very rapidly and the photon flux peaks close to the central source. On the other hand, if $\Gamma m_a r/p_a\sin\theta\lesssim 1$, the exponential suppression becomes irrelevant, and the photon flux becomes mostly independent of $\beta$ for $\beta\ll 1$. Notice that this is still much more peaked toward small angles $\beta$ than an isotropic flux, for which we would have $d\Phi_\gamma/dE_\gamma d\beta \propto \sin\beta$, but is spread over a wider angular range than the usual point source predictions. Thus, the best region for constraints is close to the center, but not exactly at the center where the standard astrophysical signal from M82 is largest. This motivates us to perform an angular and spectral analysis to draw our constraints.

Figure~\ref{fig:spectra} collects the spectral and angular distribution of the produced radiation for increasing axion-photon coupling $g_{a\gamma}$ and a fixed $m_a=316$~keV; we denote by $\mathcal{F}_\gamma=\int \Phi_\gamma E_\gamma dE_\gamma$ the photon energy flux. The central benchmark coupling choice corresponds to the constraint we obtain and it is shown in Fig.~\ref{fig:bounds}. Around this value of coupling, the photons from axion decay have two unique features compared to the standard M82 signal: they are spread over a range of a few arcminutes (comparable with the angular range we use for our analysis), while still being visibly nonisotropic; and they peak at energies of $100$~keV, while in the standard case hot thermal photons from stellar cores with comparable energies of course cannot reach us.

\textbf{\textit{NuSTAR data and results.}}---
We analyze and process 26 observations of M82, totaling $\sim$1.96 Ms of exposure time, obtained from archival data of the NuSTAR Observatory~\cite{NuSTAR:2013yza}. NuSTAR, the first high-energy focusing X-ray telescope in space, operates within the $3–79$ keV energy range, making it highly suitable for probing an axion signal from the stars of M82. It consists of two identical coaligned instrument lines, each equipped with independent optics and focal-plane module (FPM) detectors, referred to as FPMA and FPMB. Each telescope subtends a field of view of approximately \(13'\times13'\), with an angular resolution of about $60''$ half-power diameter (HPD) for a point source located near the optical axis. These grazing-incidence telescope modules enable NuSTAR to focus X-rays efficiently onto its focal plane detectors, providing high sensitivity and resolution for our analysis. Data from both detectors were reduced using \texttt{HEASoft} software version 6.34~\cite{ascl:1408.004} and \texttt{NuSTARDAS} version 2.1.4. We defined a circular source region with a radius of $60^{\prime\prime}$ centered at the coordinates of M82 (in Right Ascension (RA) and Declination (Dec), epoch=J2000):  RA$ = 09 ^\textrm{h} 55^\textrm{m} 52.43^\textrm{s}$, DEC $= +69^\circ 40' 46.9'' $. Additionally, we define 8 concentric annular regions ranging from $60''$ to $237''$, with an equidistant separation of $22''$ (see SupM~\cite{supplementalmaterial}). Spectra are extracted from each annular region, allowing us to investigate the angular distribution of photons. We do not perform any background subtraction, as the signal is anticipated to encompass the entire field of view (FOV), but we later include an isotropic background as a free fit parameter. As detailed in the SupM~\cite{supplementalmaterial}, we have also explored alternative background treatments (such as using an OFF region and incorporating a power-law component), which result in slightly stronger constraints, thereby confirming the conservative nature of our primary approach.
We stack and regroup the data into 5-keV wide energy bins, focusing on a fiducial energy range of $30–70$ keV. We do not consider energies below 30 keV to mitigate potential effects of insufficiently accurate modeling of low-energy astrophysical X-ray emission from the galaxy, following the approach outlined in \cite{Ning:2024eky}. To compute the expected axion decay signal counts at the NuSTAR FPMs for a given set of nuisance parameters $\boldsymbol{\theta} = \{g_{a\gamma}, m_a\}$, we forward-model the photon signal from axion decay through the instrument response files and exposure times of each observation. This process is performed for each energy bin $i$, angular bin $\alpha$, observation, and module. We then sum over all observations and both modules to obtain the total expected counts $N_{i,\alpha}^{a}(\boldsymbol{\theta})$.

In order to obtain the bounds from M82, we model the signal as the superposition of the axion decay component $N_{i,\alpha}^{a}(\boldsymbol{\theta})$ in each angular and energy bin, and an isotropic background.
 We model the latter with a completely free energy spectrum, so we treat the fluxes in each of the eight energy bins as nuisance parameters, and the count rate $N_{i , \alpha}^{\rm bg}(\boldsymbol{\theta})$ in each angular bin is proportional to the solid angle of the bin. We do not include here an energy-dependent component for the astrophysical signal in the central bin, which is often modeled as a power law; this is conservative, because such a signal acts as a background for astrophysical searches. More importantly, as further discussed in the SupM~\cite{supplementalmaterial}, we have explicitly verified that our constraints are only marginally increased by subtracting off the background using an OFF region and modeling the photon signal from the central M82 source with a power law. This can be explained by taking into account that for the values of the coupling we are probing the main contribution to the axion signal comes from an ON region of radius $\lesssim 2^\prime$ and, for the largest masses we are interested in [$m_a \gtrsim O(100)$~keV], constraints are dominated by the high-energy bins in the ON region, where the astrophysical background becomes smaller.
The observed counts in each bin are denoted by $N_{i \alpha}^{\rm obs}$. To obtain our bounds, we introduce a log-likelihood $\Lambda$
\begin{align}
    \Lambda(g_{a\gamma},m_a)&= \\
    \nonumber
    &\mathrm{max}_{\boldsymbol{\theta}} \sum_{i,\alpha}\left[N_{i \alpha}^{\rm obs}\log (N_{i\alpha}^a+N_{i\alpha}^{\rm bg})\right.
    \left.
    -N_{i\alpha}^a-N_{i\alpha}^{\rm bg}\right].
\end{align}
We now define a test statistic (TS) following the prescriptions of Ref.~\cite{Cowan:2010js} as 
${\rm TS}=-2(\Lambda-\mathrm{max}_{g_{a\gamma}}\Lambda)$ if $g_{a\gamma}$ is above the best-fit value that maximizes $\Lambda$, and ${\rm TS}=0$ otherwise. With this definition, the TS follows a half-chi-squared distribution with one degree of freedom, and the upper bounds on $g_{a\gamma}$ can be obtained by the threshold condition TS$=2.7$.

\textbf{\textit{Discussion and outlook.}}---Our bounds, shown in Fig.~\ref{fig:bounds}, are more than an order of magnitude stronger than other astrophysical bounds in the mass range of $30-300$~keV. Part of the excluded region is already ruled out by arguments based on the freeze-in of the axions in the early Universe; this is a cosmological observable highly complementary to the astrophysical one introduced here. An additional cosmological observable that can rule out parameter space of interest to us is the spectral distortion (SD) of the cosmic microwave background (CMB) induced by axions decaying to photons in the early Universe. Here we report these bounds as shown in Ref.~\cite{Langhoff:2022bij}, which have been indirectly obtained from Ref.~\cite{Balazs:2022tjl}. Our arguments rule out a previously uncharted region of parameter space which, at masses of several hundreds of keV, supersedes the previous bounds by nearly an order of magnitude.

NuSTAR provides an excellent instrument for the signature we identified, due to its angular resolution that allows us to probe the angular distribution over angles close to a few arcminutes from the source. Notice also that the constrained couplings lead to a decay length larger than $\sim\,\rm Mpc$; that means that extragalactic sources are best suited for this kind of searches. On the other hand, Fig.~\ref{fig:spectra} also shows that the energy window probed by NuSTAR is not directly centered on the peak of the photon spectrum. Therefore, measurements at higher photon energies in the hundreds of keV range, corresponding to the range of, e.g., INTEGRAL, could offer some advantage and complementarity to the strategy followed here. This is especially true since a tiny portion of the axions, produced with nonrelativistic velocities below the escape velocity of their progenitor star, might remain trapped around the star and form a ``basin'' similar to the solar basin. These axions, being at rest, would produce photons in the hundreds of keV region. We will consider the potential of these complementary signatures in a future work.

We should also distinguish our new observable from the recent proposal of Ref.~\cite{Chen:2024ekh}, which considers a scenario in which axions are produced in the stellar cores through electron and photon couplings within Alpha Centauri, remain gravitationally trapped, and subsequently decay nearly at rest into photons. In our case, the temperatures of the stars dominating axion production are much higher, so our bounds reach up to much larger axion masses.

Notice that in this work we assume a minimal axion coupling to photons only. Obviously, for a specific UV-complete model, couplings to other species could be present as well. We focus here on the introduction of a novel observable whose relevance is general; the specifics of a given model only affect the details, but not the core, of our approach. For example, an additional coupling to electrons could enhance the axion emissivity from stellar cores, leading to strong constraints on the combined parameter space; this would only enlarge the parameter space without providing new insight. On the other hand, for a given UV-complete model, our approach can easily be extended to include the correct production processes.

To summarize, we have found that the cores of heavy stars within SBGs are by far the most powerful astrophysical probe of axion production in the tens-to-hundreds of keV mass range. Cosmological observables offer orthogonal probes, but our new argument can still exclude regions of parameter space that were previously allowed. By the same token, we expect other FIPs with radiative decay channels, such as sterile neutrinos, to also be probed by similar arguments. In this way, we have extended the observability of radiatively decaying FIPs to a new class of astrophysical sources.

\textbf{\textit{Acknowledgments}}---We warmly thank Sebastian Hoof for discussions in the early stages of the project, Nicholas Rodd for discussions about the axion cosmological bounds and suggestions on a draft of this paper, and Georg Raffelt for important comments. This work was internally reviewed by Chengchao Yuan. FRC is supported by the Universidad de Zaragoza under the ``Programa Investigo'' (Programa Investigo-095-28), as part of the Plan de Recuperación, Transformación y Resiliencia, funded by the Servicio Público de Empleo Estatal and the European Union-NextGenerationEU. DFGF is supported by the Alexander von Humboldt Foundation (Germany) and, when this work was begun, was supported by the Villum Fonden (Denmark) under Project No. 29388 and the European Union’s Horizon 2020 Research and Innovation Program under the Marie Sklodowska-Curie Grant Agreement No. 847523 ``INTERACTIONS.''. GL acknowledges support from the U.S. Department of Energy under contract number DE-AC02-76SF00515 and, when this work was started, the EU for support via ITN HIDDEN (No 860881). EV is supported by the Italian MUR Departments of Excellence grant 2023-2027 ``Quantum Frontiers'' and by Istituto Nazionale di Fisica Nucleare (INFN) through the Theoretical Astroparticle Physics (TAsP) project. This article is based upon work from COST Action COSMIC WISPers CA21106, supported by COST (European Cooperation in Science and Technology)

\bibliographystyle{bibi}
\bibliography{References}
\onecolumngrid
\appendix

\setcounter{equation}{0}
\setcounter{figure}{0}
\setcounter{table}{0}
\setcounter{page}{1}
\makeatletter
\renewcommand{\theequation}{S\arabic{equation}}
\renewcommand{\thefigure}{S\arabic{figure}}
\renewcommand{\thepage}{S\arabic{page}}

\begin{center}
\textbf{\large Supplemental Material for the Letter\\[0.5ex]
{\em NuSTAR bounds on radiatively decaying particles from M82 
}}
\end{center}

In this Supplemental Material, we detail our modeling of the axion production in M82, of the photon angular and spectral distribution from the axion decay, and our analysis of the data from the Nuclear Spectroscopic Telescope Array (NuSTAR) mission. We also show our constraints over an extended region of mass and coupling, to provide a broader comparison with the pre-existing bounds.

\bigskip

\section{A.~Axion production in M82}
In the stellar plasma, axions interacting with photons may be produced via the Primakoff process on charged particles $\gamma + Ze \to Ze + a$ and photon coalescence $\gamma \gamma \to a$.

The Primakoff production rate for massive axions is~\cite{DiLella:2000dn,Lucente:2020whw,Caputo:2022mah}
\begin{equation}
\Gamma_{\rm P}=\,g_{a\gamma}^2\dfrac{T\kappa_s^2}{32\pi} \dfrac{p_a}{E_a}\left\{\dfrac{\left[\left(k_\gamma+p_a\right)^2+\kappa_s^2\right]\left[\left(k_\gamma-p_a\right)^2+\kappa_s^2\right]}{4k_\gamma p_a\kappa_s^2} \ln\left[\dfrac{(k_\gamma+p_a)^2+\kappa_s^2}{(k_\gamma-p_a)^2+\kappa_s^2}\right]
-\dfrac{\left(k_\gamma^2-p_a^2\right)^2}{4k_\gamma p_a\kappa_s^2}\ln\left[\dfrac{(k_\gamma+p_a)^2}{(k_\gamma-p_a)^2}\right]-1\right\}\,, \\*
\label{eq:Prima}
\end{equation}
where $T$ is the temperature, $p_a = \sqrt{E_a^2 - m_a^2}$ and $k_\gamma = \sqrt{E_a^2 - \omega_p^2}$ are the axion and photon momenta, ${\omega_p^2= 4\pi\alpha \sum_i Z_i^2 n_i/m_i}$ is the plasma frequency and $\kappa_s^2=\frac{4\pi\alpha}{T}\sum Z_i^2 n_i$ the Debye screening scale. Here, $\alpha$ is the electromagnetic fine structure constant, and the sum runs over species $i$ with electric charge $Z_i e$, mass $m_i$ and number density $n_i$.
By convolving the production rate with the photon density we obtain the axion production spectrum per unit volume in units of keV$^{-1}$~s$^{-1}$~cm$^{-3}$,

\begin{equation}
\begin{split}
    \left(\frac{d\dot{n}_{a}}{dE_a}\right)_{\rm P}=\,g_{a\gamma}^2\dfrac{T\kappa_s^2\,p\,E_a}{32\pi^3} \dfrac{1}{e^{E_a/T}-1}\bigg\{&\dfrac{\left[\left(k_\gamma+p_a\right)^2+\kappa_s^2\right]\left[\left(k_\gamma-p_a\right)^2+\kappa_s^2\right]}{4k_\gamma p_a\kappa_s^2} \ln\left[\dfrac{(k_\gamma+p_a)^2+\kappa_s^2}{(k_\gamma-p_a)^2+\kappa_s^2}\right]-\\
    &-\dfrac{\left(k_\gamma^2-p_a^2\right)^2}{4k_\gamma p_a\kappa_s^2}\ln\left[\dfrac{(k_\gamma+p_a)^2}{(k_\gamma-p_a)^2}\right]-1\bigg\}\,. 
\end{split}
\end{equation}

For $m_a \gtrsim T$, axions can be efficiently produced via photon coalescence $\gamma\gamma \to a$, which can be seen as the reverse process of axion decay. Taking into account the Bose-Einstein stimulation of the final-state photons and detailed balance, the production spectrum per unit volume for this process can be computed as~\cite{Caputo:2022mah}
\begin{equation}
    \left(\frac{d\dot{n}_{a}}{dE_a}\right)_{\rm C}=\,\frac{g_{a\gamma}^2\,T\,m_a^4}{64 \pi^3} \frac{e^{-E_a/T}}{1-e^{-E_a/T}} \log\left[\frac{e^{\frac{E_a+p_a}{4T}}-e^{-\frac{E_a+p_a}{4T}}}{e^{\frac{E_a-p_a}{4T}}-e^{-\frac{E_a-p_a}{4T}}}\right]\,,
\end{equation}
For $m_a \gg T$ this result agrees with the expressions in Ref.~\cite{DiLella:2000dn,Lucente:2020whw,Carenza:2020zil} obtained using Maxwell-Boltzmann statistics. 

\begin{figure}[t!]
    \includegraphics[width=0.49\textwidth]{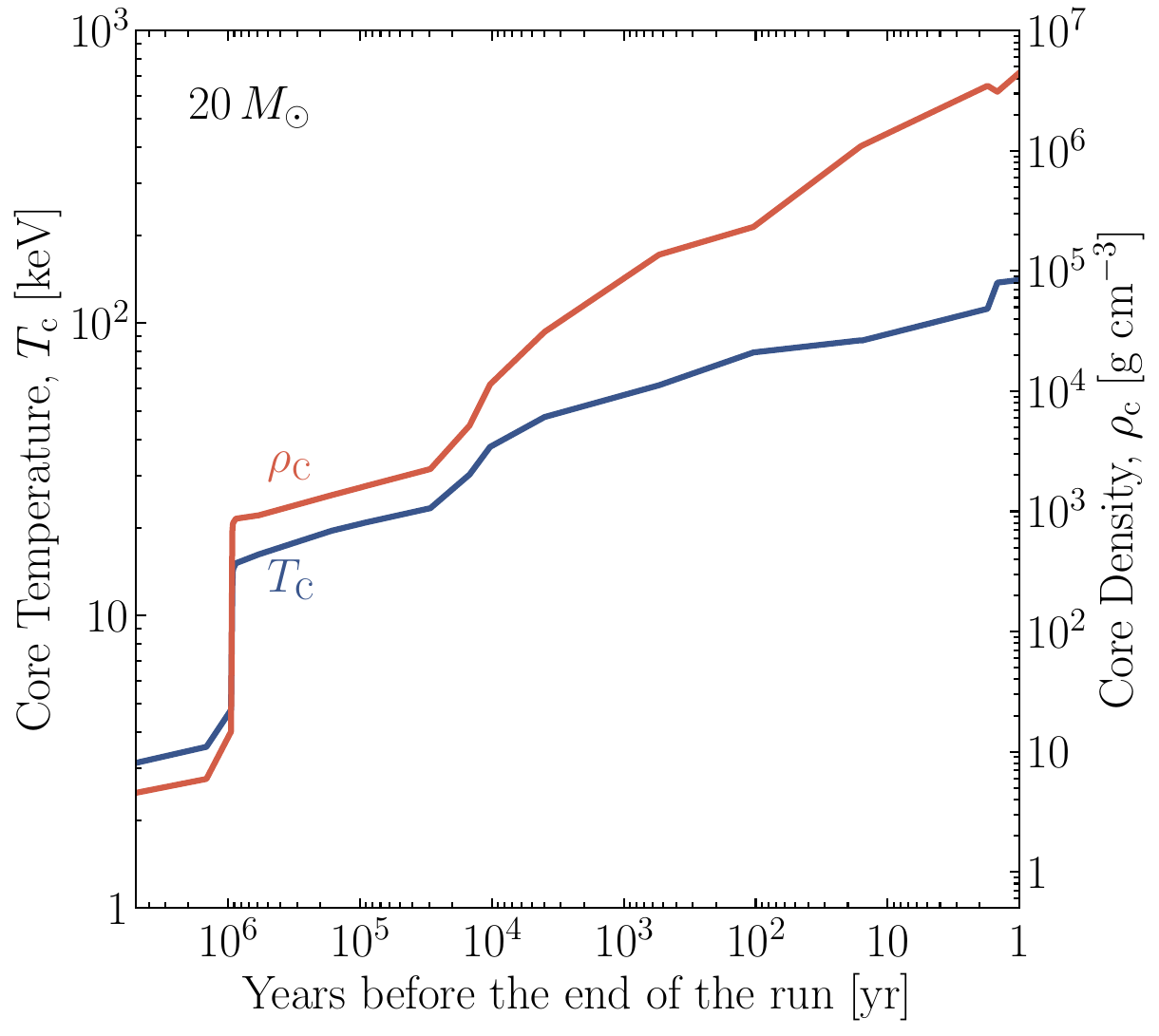}
    \includegraphics[width=0.49\textwidth]{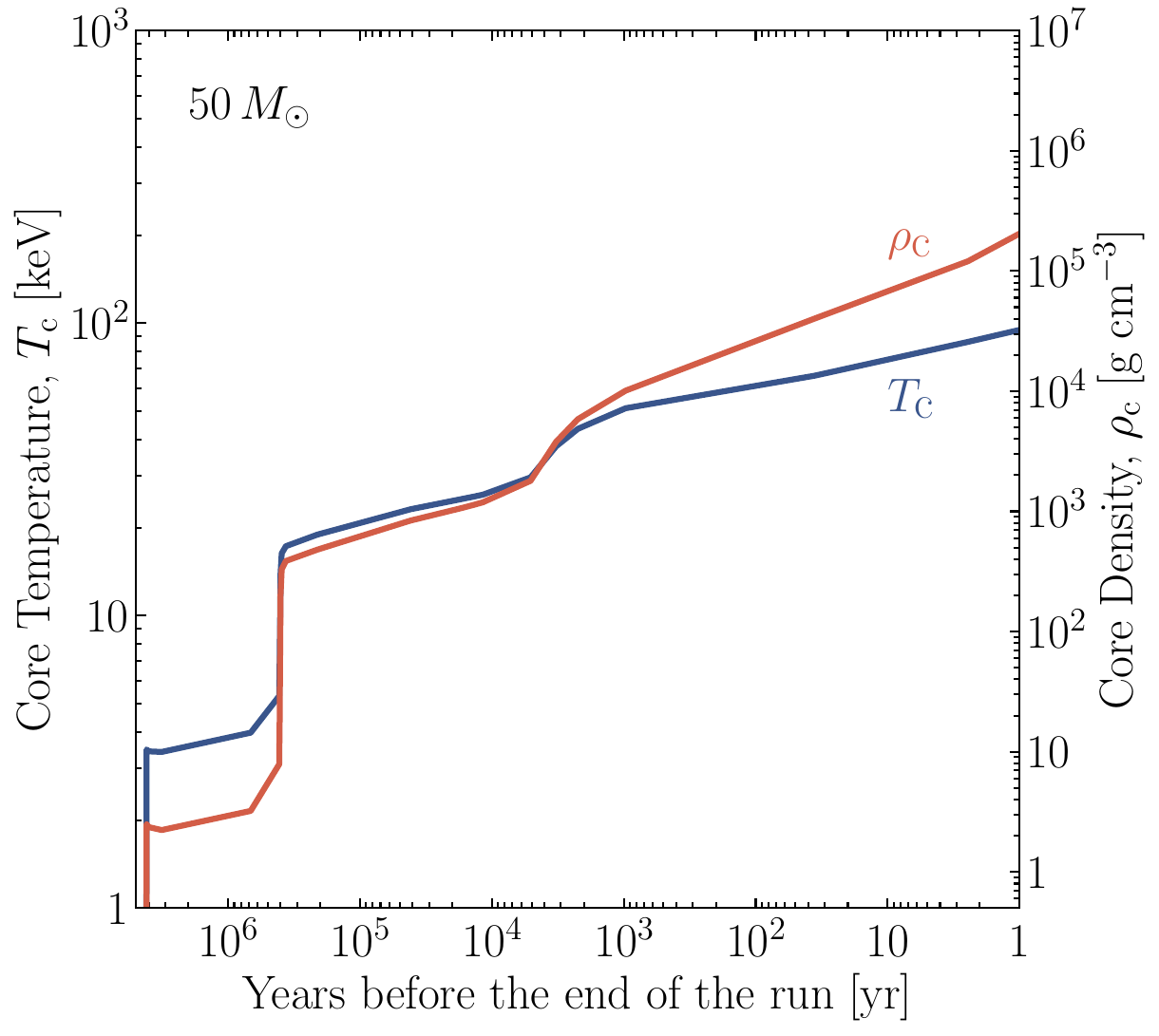}
    \caption{Time evolution of the central core temperature (blue) and density (red) for an initial $20~M_\odot$ (left) and $50~M_\odot$ (right) star.}\label{fig:profiles}
\end{figure}

By integrating $d\dot{n}_a/dE_a$ over the stellar profile, we compute the axion production spectrum from a single star $d\dot{N}_a^s/dE_a$. We integrate the production spectra per unit volume over stellar profiles obtained through the one-dimensional stellar evolution code Modules for Experiments in Stellar Astrophysics (MESA)~\cite{Paxton:2010ji,Paxton:2013pj} (release {\tt r23.05.1}). Following Ref.~\cite{Ning:2024eky}, we evolve stellar models with different initial masses and a metallicity of $Z=0.02$, based on spectral data from Ref.~\cite{Origlia:2004tv}. As further discussed in Ref.~\cite{Ning:2024eky}, the dominant contribution to axion emission in M82 comes from high-mass stars (with mass $M\gtrsim 20~M_\odot$). Thus, here we use a grid of $10$-$M_\odot$-spaced models between $20~M_\odot$ and $80~M_\odot$, evolved using the default MESA suite and in-list for high-mass stars, {\tt 20M\_pre\_ms\_to\_core\_collapse}, from pre-main sequence to the onset of core-collapse. In this way we obtain radial profiles of quantities like the temperature $T$ and the density $\rho$ at different stellar ages, depending on the initial stellar mass. Massive stars with $M\gtrsim 8~M_\odot$ spend the major part of their life on the Main Sequence (MS), where hydrogen burns into helium. When the hydrogen in the core runs out, the star leaves the MS and expands, becoming a supergiant. Due to the large star mass, helium fusion starts before the core becomes degenerate. After helium exhaustion in the core, stars begin to burn heavier elements until an iron core is formed and they undergo core-collapse, resulting in a supernova explosion. The duration of the post-MS phases is much shorter than the MS. For the stellar masses we are interested in, the helium core-burning phase lasts $\sim 0.5-1$~Myr (the heavier the star, the shorter the duration), the carbon-burning one about $10^4$ yrs and later stages are even shorter, with the silicon-burning phase only few days. We show in Fig.~\ref{fig:profiles} the time evolution of the core temperature (blue) and density (red) during the late stages of an initial $20~M_\odot$ (left) and $50~M_\odot$ (right) star. The most relevant phase for our analysis is the helium burning phase between $\sim 10^6$ and $\sim 10^4$ years before the collapse, when the core temperature is $T_{\rm C}\sim \mathcal{O}(10)$~keV and the core density is $\rho_{\rm C}\sim \mathcal{O}(10^3)$~g~cm$^{-3}$. At later stages, the temperature and density increase but on much shorter timescales, therefore leading to a smaller contribution to the axion production.

\begin{figure}[t!]
    \includegraphics[width=0.47\textwidth]{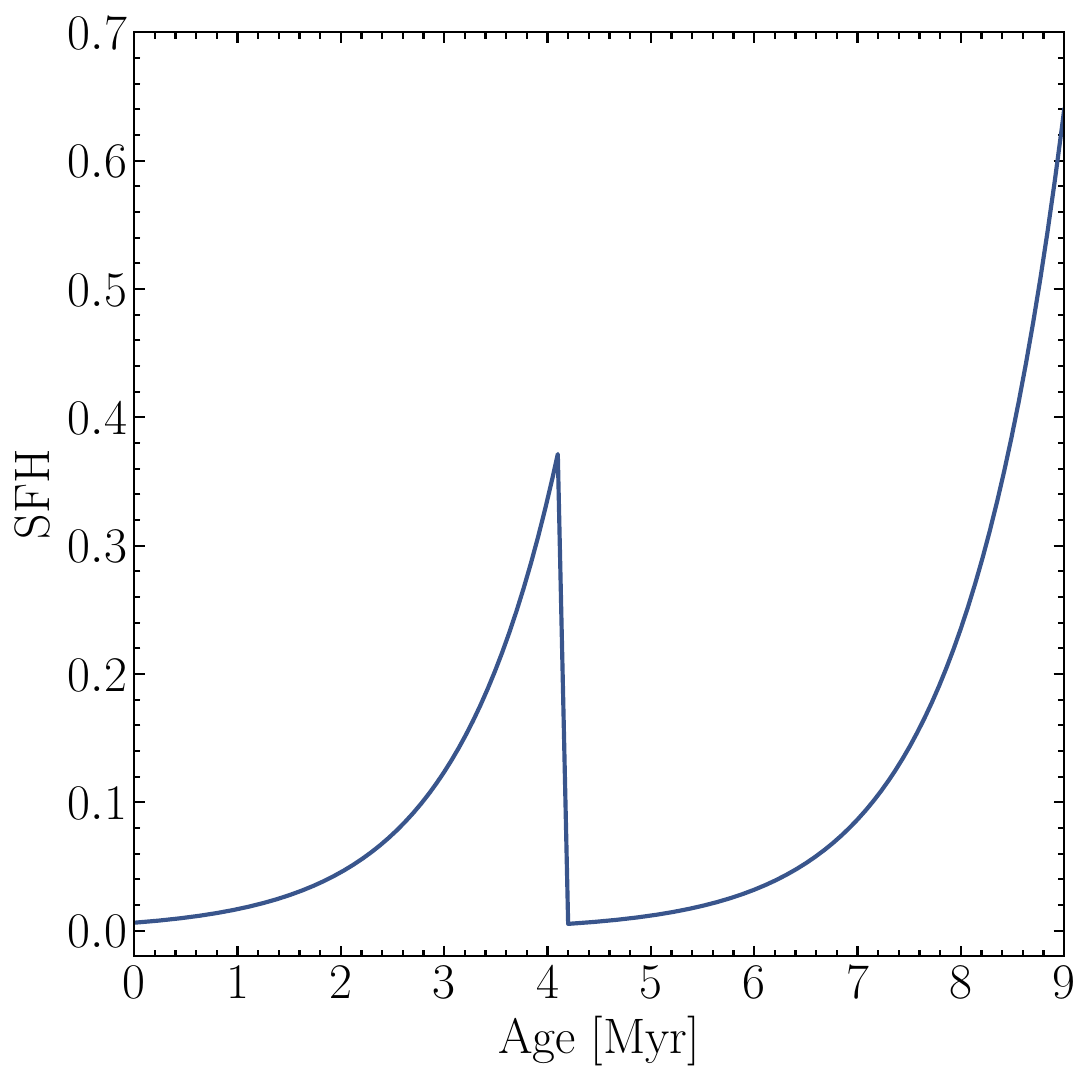}
    \includegraphics[width=0.49\textwidth]{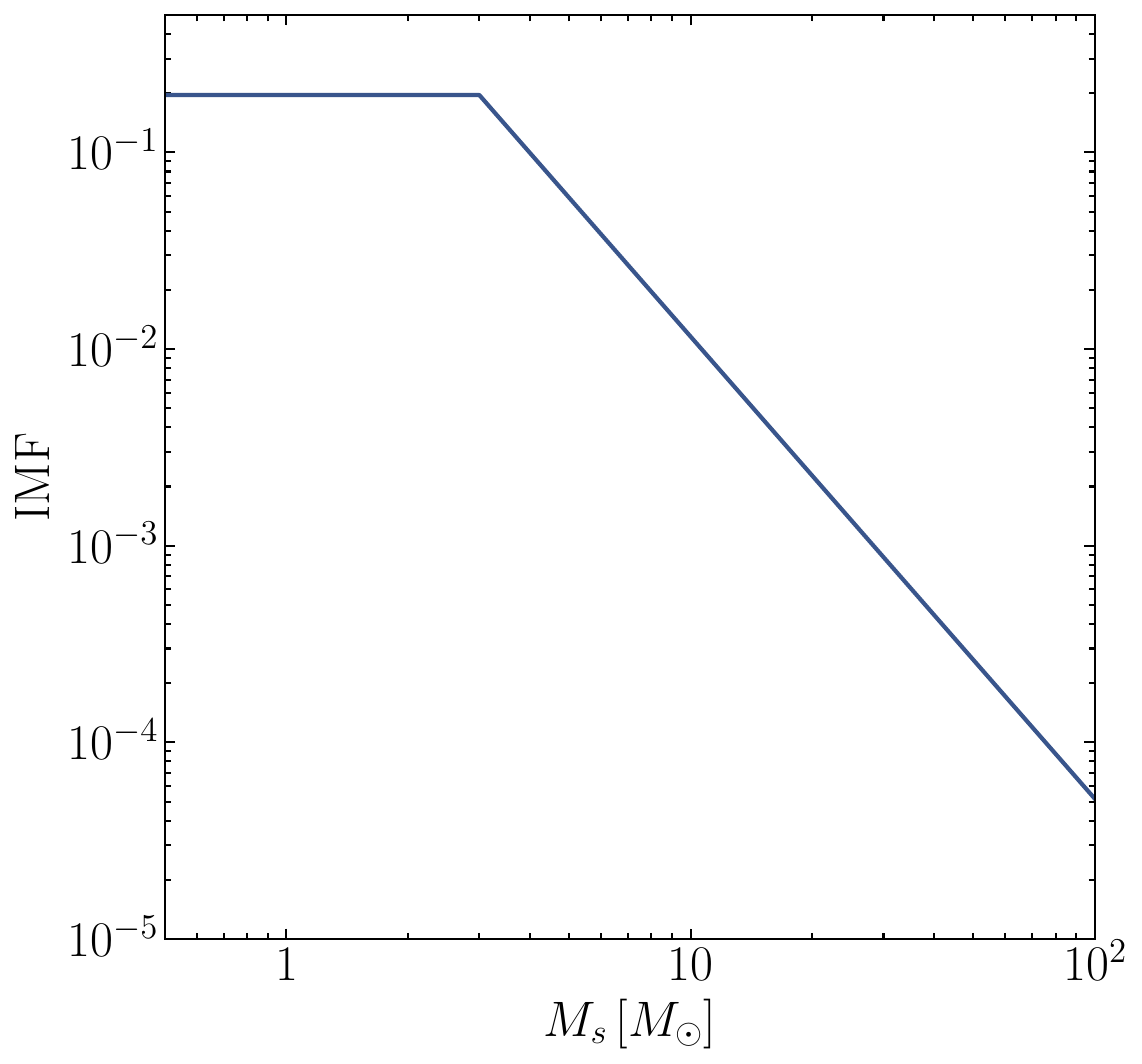}
    \caption{The SFH (left) and the IMF (right) shown as probability density functions.}\label{fig:SFH-IMF}
\end{figure}

In order to compute the axion production in M82 we need to model the stellar population through the Stellar Formation History (SFH) and the Initial Mass Function (IMF). The SFH describes the stellar distribution according to their age, quantifying how rapidly stars form in a galaxy over time. On the other hand, the IMF represents the distribution of stellar masses at the time of their formation within a given population of stars. For both the cases we follow the models used in Ref.~\cite{Ning:2024eky}. We show in the left panel of Fig.~\ref{fig:SFH-IMF} the SFH for M82. It is a ``two-burst'' model described by $R_0\,e^{(t_{\rm burst}-t)/t_{\rm sc}}$ for $t < t_{\rm burst}$, where $t_{\rm sc}=1.0$~Myr is the characteristic decay time scale. For the old burst, $t_{\rm burst}=9.0$~Myr and $R_0 = 31$~$M_\odot/{\rm yr}$, while for the younger burst $t_{\rm burst}=4.1$~Myr and $R_0 = 18$~$M_\odot/{\rm yr}$~\cite{ForsterSchreiber:2003ft}. The right panel of Fig.~\ref{fig:SFH-IMF} shows the IMF for M82, which is $\propto M^{-2.35}$ at high mass and flat below $\sim 3M_\odot$, with a cut-off at 100~$M_\odot$, as determined by comparing infrared spectroscopy data with population synthesis models~\cite{ForsterSchreiber:2003ft}.

\begin{figure}[t!]
    \includegraphics[width=0.49\textwidth]{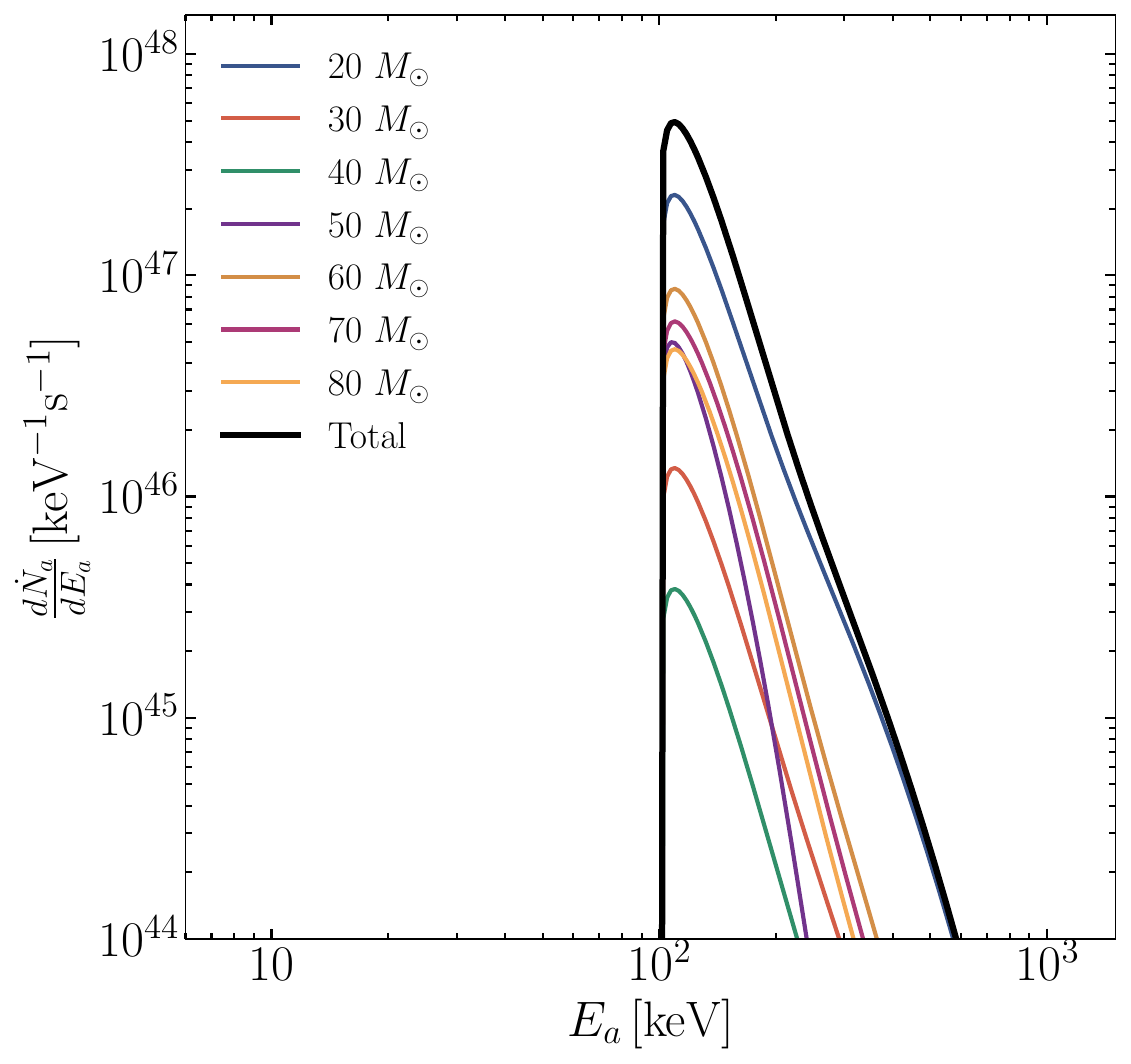}
    \includegraphics[width=0.48\textwidth]{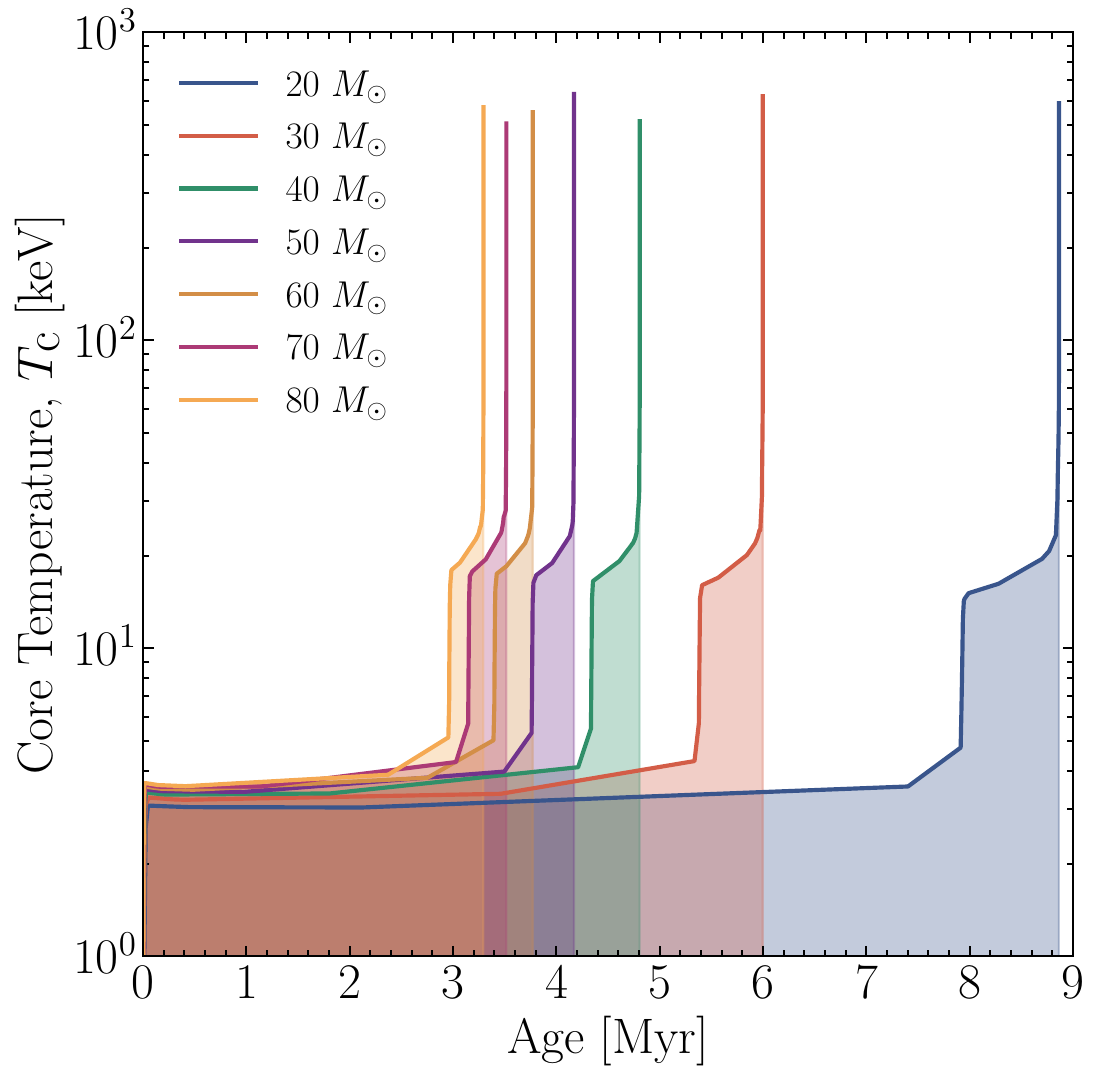}
    \caption{\emph{Left panel}: Axion production spectrum for $m_a = 100$~keV and $g_{a\gamma}=10^{12}$~GeV$^{-1}$ from the M82 galaxy (black) and from populations represented by the models with different initial mass, as shown in the legend. For the partial contribution we show $\frac{d\dot{N}_a^s}{dE_a dM_s} \Delta M_s$. \emph{Right panel}: Time evolution of the core temperature $T_{\rm c}$ at the late stages of the stellar life.}
    \label{fig:prodtempSM}
\end{figure}

Therefore, we can use the IMF and the SFH as probability density functions and compute the total axion production spectrum in M82 as
\begin{equation}
  \frac{d\dot{N}_a}{dE_a} =   N_{\rm tot} \int dM_s\, dt_s\, {\rm IMF}(M_s)\,{\rm SFR}(t_s) \,  \frac{d\dot{N}^s_{a}}{dE_a}, 
\end{equation}
where $N_{\rm tot}=1.8\times 10^{10}$~\cite{Ning:2024eky} is the total number of stars in M82 estimated using luminosity observations~\cite{NED} and color-mass-to-light ratio relations for disk galaxies from Ref.~\cite{1407.1839}, and $d\dot{N}^s_{a}/dE_a$ is the total production spectrum from a single star with age $t_s$ and initial mass $M_s$. We discretize the integration using profiles from our grid of $10~M_\odot$ spaced models between $20~M_\odot$ and $80~M_\odot$ and we find that the dominant contribution to the axion production in M82 comes from the population represented by $20~M_\odot$ initial mass stars, as shown in the left panel of Fig.~\ref{fig:prodtempSM} for $m_a=100$~keV and $g_{a\gamma}=10^{-12}$~GeV$^{-1}$. Indeed, the dominant contribution to the axion production in M82 comes from supergiant stars with the core temperature $T_{\rm C}\sim \mathcal{O}(10)$~keV. As shown in the right panel of Fig.~\ref{fig:prodtempSM}, the core temperature reaches a plateau at $T_{\rm C}\sim \mathcal{O}(10)$~keV for $0.5-1$~Myrs (the heavier the initial mass, the shorter this phase and the higher the core temperature), before starting to increase when elements heavier than helium are burnt on much shorter timescales $\lesssim 10^4$~yrs. The core temperature plateau for the $20~M_\odot$ star is located at a stellar age $\sim 8-9$ Myrs, where the SFH for M82 is peaked (see the left panel of Fig.~\ref{fig:SFH-IMF}). A non-negligible contribution comes also from stars with initial mass $50-60~M_\odot$, since the temperature plateau is at an age $3.5-4$~Myrs, where the peak of the younger burst in the SFH takes place. Stars with initial mass $M\lesssim 20~M_\odot$ remain in the main sequence phase for longer time and therefore they become supergiants with $T_{\rm C}\sim \mathcal{O}(10)$~keV later (e.g. an initial $10~M_\odot$ star is on the MS for $\sim 20$~Myrs~\cite{1996ima..book.....C}), implying that their contribution to the axion production is negligible.

\begin{figure}[t!]
    \includegraphics[width=0.49\textwidth]{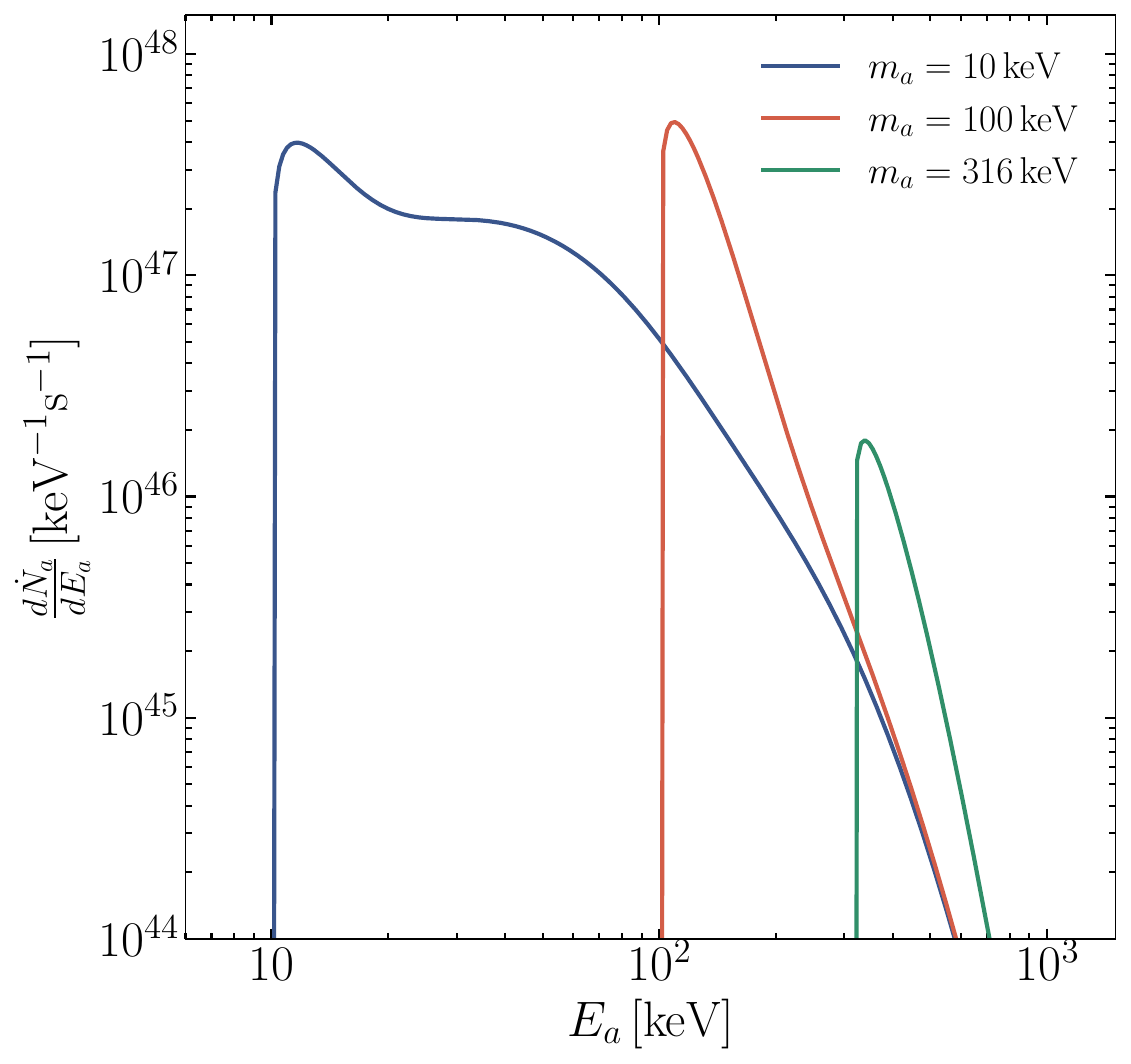}
    \includegraphics[width=0.49\textwidth]{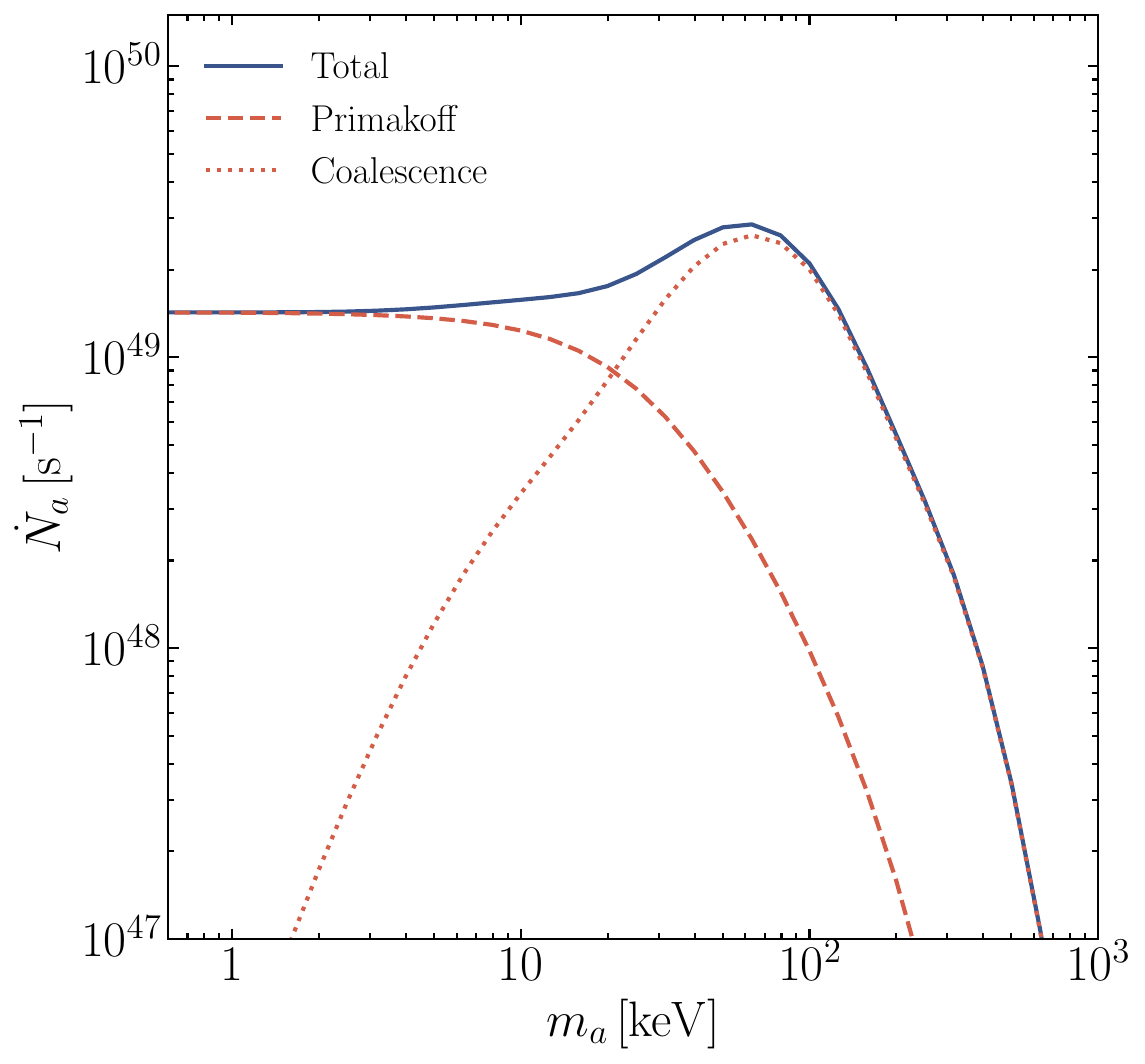}
    \caption{\emph{Left panel}: Total axion production spectra from M82 for $m_a=10$~keV (blue), $m_a=100$~keV (red), $m_a=316$~keV (green). \emph{Right panel}: Axion production rate from M82 via Primakoff (dashed red), photon coalescence (dotted red) and including both the processes (solid blue). In both the panel we use $g_{a\gamma}=10^{-12}$~GeV$^{-1}$.}\label{fig:spectraSM}
\end{figure}

We show in the left panel of Fig.~\ref{fig:spectraSM} the total production spectra in M82, including both the Primakoff and the photon coalescence contributions, for $g_{a\gamma}=10^{-12}$~GeV$^{-1}$ and three different masses $m_a = 10$~keV in blue, $m_a = 100$~keV in red and $m_a = 316$~keV in green. At $m_a=10$~keV both the processes are relevant, with the coalescence leading to the bump for $10~{\rm keV} \lesssim E_a \lesssim 20$~keV and the Primakoff dominant at larger energies. On the other hand, spectra at $m_a=100$~keV and $m_a=316$~keV are mainly determined by photon coalescence. 
The right panel of Fig.~\ref{fig:spectraSM} shows the number of axions produced per unit time in M82 as a function of the axion mass, obtained integrating over energies the production spectra. As expected, since the dominant contribution to the axion production comes from stars with $T_{\rm C}\sim \mathcal{O}(10)$~keV, the total production rate (solid blue) is peaked at $m_a \simeq 60$~keV before being Boltzmann-suppressed. For comparison, we show also the production rate via Primakoff (dashed red line) and coalescence (dotted red). The photon coalescence contribution becomes dominant at $m_a \gtrsim 30$~keV, before being Boltzmann suppressed. This is in agreement with the results obtained in the context of supernova axions (see Refs.~\cite{Lucente:2020whw,Caputo:2022mah}), where $T\simeq 30$~MeV and the coalescence is dominant at $m_a\gtrsim 60$~MeV. Indeed, the most relevant parameter for the axion production processes is the temperature $T$, with the density $\rho$ contributing only with $\mathcal{O}(1)$ correction factors.

\section{B.~Axion decay}

Axions are produced by an essentially steady source, namely the reservoir of stars in the starburst galaxy M82, and their subsequent decay produces photons. In this sense, the mechanism is quite similar to the analogous expected signal from the decay of heavy particles produced in the core of SN~1987A. However, for the latter case the spectrum of the daughter particles is usually obtained under the assumption that the heavy particle decays very close to the supernova, at distances much shorter than the distance from the Earth. For the case of neutrino-decaying bosons, this condition is typically ensured for the couplings that can be constrained by the non-observation of high-energy neutrinos. For the case of particles decaying to photons, in principle the much better sensitivity to an electromagnetic signal would allow one to probe even couplings so low that this condition would be mildly violated. However, because the Solar Maximum Mission (SMM) only observed the signal for 232.2~s, after which it went into calibration mode for 10~minutes. The photons produced within this early time would have come from heavy particles decaying very close to the supernova anyway, so that the assumption is still valid for the relevant part of the signal.

\begin{figure*}[t!]
    \includegraphics[width=\textwidth]{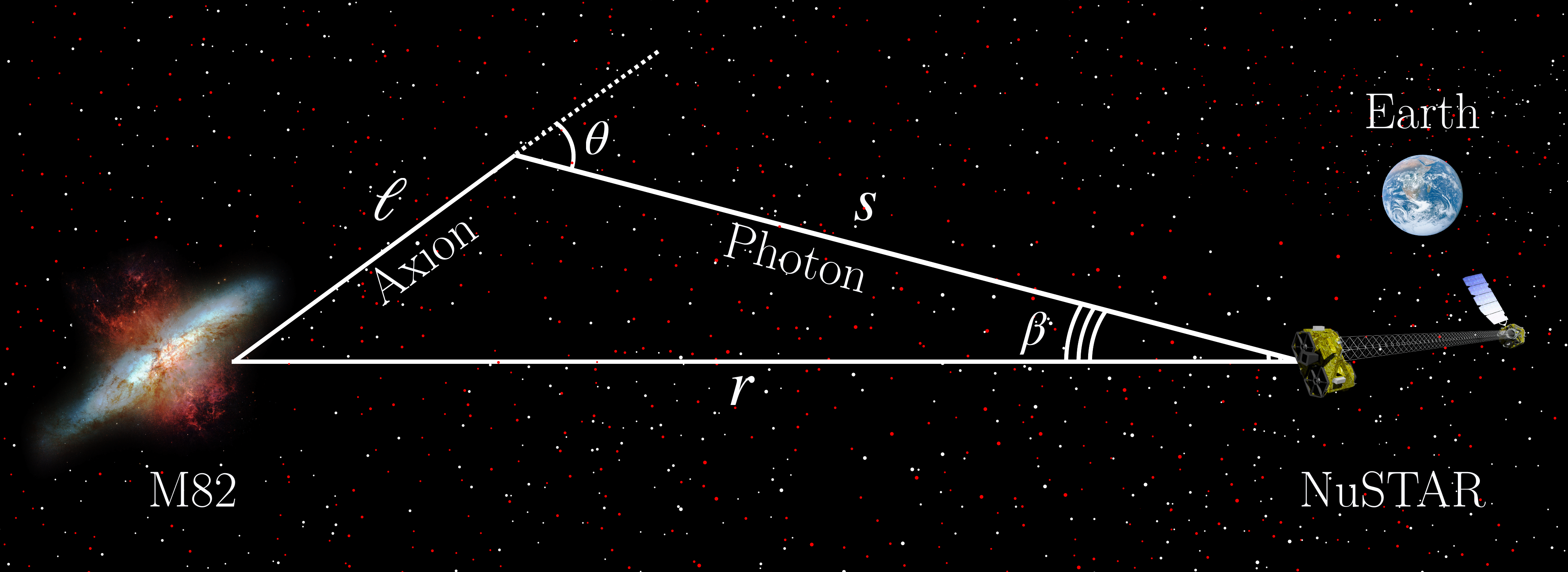}
    \caption{Sketch of the axion decay into photons, together with the notation that we use for the calculation of the photon signal.}\label{fig:decay_sketch}
\end{figure*}

In our case, however, this assumption is untenable. M82 is a steady source, so if the decay length becomes comparable with the distance between M82 and the Earth, a sizable portion of the signal will always come from axions decaying far from M82. The key observation is that for these axions, despite the potentially large Lorentz boosting, the photon signal will arise at a relatively large angle from the nucleus of M82.
In this section, we derive the corresponding energy and angular distribution. We use the geometry sketched in Fig.~\ref{fig:decay_sketch}. Compared to the derivations in Ref.~\cite{Oberauer:1993yr}, we are going to relax the approximation $\ell\ll r$. In principle, we proceed with the derivation also for generic values of the angle $\beta$, although in the end we will only be interested in the regime $\beta\ll 1$.

We assume that axions are produced from the point source with an emission rate $d\dot{N}_a/dE_a$ in the radial direction, with $E_a$ and $m_a$ the energy and mass of the axion, and decay with a rest-frame rate $\Gamma_a$. The notation we use to determine the photon flux follows the sketch in Fig.~\ref{fig:decay_sketch}. Thus, the number of axions per unit volume at a distance $\ell$ from the source is
\begin{equation}
    \frac{dN_a}{dVdE_a}=\frac{d\dot{N}_a/dE_a}{4\pi\ell^2 v_a}\exp\left[-\frac{\Gamma_a m_a\ell}{v_a E_a}\right],
\end{equation}
where $v_a=\sqrt{1-m_a^2/E_a^2}$ is the velocity of the axion.

The decay of an axion produces two photons, which in the rest frame of the axion have an energy $m_a/2$. In the laboratory frame, the energy distribution of the produced photons is
\begin{equation}
    P(E_\gamma)dE_\gamma=\frac{dE_\gamma}{p_a}\Theta\left(\frac{E_a+p_a}{2}-E_\gamma\right)\Theta\left(E_\gamma-\frac{E_a-p_a}{2}\right),
\end{equation}
where $p_a=E_a v_a$ is the momentum of the axion, and $\Theta$ denotes the Heaviside function. Since the axion moves radially, the angle with which the photon is emitted from the radial direction is 
\begin{equation}\label{eq:costheta}
    \cos\theta=\frac{1}{v_a}\left[1-\frac{E_a}{2E_\gamma}(1-v_a^2)\right].
\end{equation}
We can now write a kinetic equation for the distribution functions of the photons $f_\gamma(r,\cos\beta,E_\gamma)$, where $\beta$ is the angle between the photon and the radial direction~\footnote{We use this somewhat unconventional notation for the polar angle to match the geometry of Fig.~\ref{fig:decay_sketch}, where $\beta$ is the polar angle at the detection point, while $\theta$ is the polar angle at the decay point.} 
\begin{equation}
    \cos\beta\frac{\partial f_\gamma}{\partial r}+\frac{\sin^2\beta}{r}\frac{\partial f_\gamma}{\partial \cos\beta}=\int_{E_{a,\mathrm{min}}}^{+\infty}\frac{2dE_a}{p_a}\frac{d\dot{N}_a/dE_a}{4\pi r^2 v_a}\exp\left[-\frac{\Gamma_a m_ar}{v_a E_a}\right]\frac{\Gamma_a m_a}{E_a} \delta\left(\cos\beta-\cos\theta\right);
\end{equation}
the minimum energy $E_{a,\mathrm{min}}=E_\gamma+m_a^2/4E_\gamma$ is obtained from the kinematic conditions, the factor $2$ accounts for the two photons produced in the decay, the $\delta$ function gives their angular distribution immediately after the decay, and the factor $\Gamma_a m_a/E_a$ gives the decay rate of the axions. 
In detail, $f_\gamma$ is defined as the differential number of photons $f_\gamma=dN_\gamma/dE_\gamma d\cos\beta$.

This equation must be integrated along the trajectories of unperturbed photons, defined by the condition $r\sin\beta=\mathrm{const.}$, with the condition that at $r\to\infty$ there should be no ingoing particles with $\cos\beta\to -1$. The integral of the $\delta$ function gives an Heaviside theta function, 
so that we are finally led to the solution
\begin{equation}\label{eq:photon_distribution}
    f_\gamma=\int_{E_{a,\mathrm{min}}}^{+\infty}\frac{2dE_a}{p_a}\frac{d\dot{N}_a/dE_a}{4\pi r \sin\beta}\Theta(\cos\beta-\cos\theta) \exp\left[-\frac{\Gamma_a m_a r \sin\beta}{p_a\sin\theta}\right]\frac{\Gamma_a m_a}{p_a\sin\theta};
\end{equation}
notice that the condition enforced by the $\delta$, leading to the $\Theta(\cos\beta-\cos\theta)$, simply leads to the geometrically obvious constraint that $\beta<\theta$. This is the expression we report in the main text, and that we use for all of our computations. Still, for qualitative understanding, it is useful too consider explicitly the approximate dependence of the angular distribution in terms of the axion parameters.

The regime that is most interesting for us is at heavy masses, $m_a>100$~keV. Since the typical gamma-ray energies we are interested in are below $E_\gamma\lesssim 80$~keV, we can adopt the approximation $m_a\gg E_\gamma$. In this approximation, the minimal axion energy for producing the photon is $E_{a,\rm min}\simeq m_a^2/4E_\gamma\gg m_a$. In this regime, since the typical energy of the axion spectrum peaks below $100$~keV, the spectrum $d\dot{N}_a/dE_a$ drops rapidly with the axion energy, so that the integral is entirely dominated by the lower bound $E_a\sim E_{a,\rm min}$. However, extremely close to the lower bound the angular-dependent term is close to 0, because $\sin\theta\to 0$. The physical reason for this is that the massive axions have energies much larger than the photons in the tens of keV range. Thus, in order for such an energetic axion to produce a low-energy photon, the decay must essentially be backwards, with $\cos\theta\sim -1$.
Therefore, we can write the typical energy of the axions contributing to the gamma-ray signal as $E_a\simeq \frac{m_a^2}{4E_\gamma}+E_\gamma \eta^2$, where $\eta$ is a dimensionless parameter. 

Close to the threshold, we approximate 
\begin{equation}
    \frac{d\dot{N}_a}{dE_a}(E_a)\simeq \frac{d\dot{N}_a}{dE_a}\left(\frac{m_a^2}{4E_\gamma}\right)\exp\left[-\frac{E_\gamma}{T}\eta^2\right],
\end{equation}
where $T$ is an effective temperature parameterizing the expected exponential decrease at high energies; we may expect $T\sim 100$~keV. The entire scheme of approximation requires $\eta\ll m_a/2 E_\gamma$. With these approximations, the integral can be expanded as
\begin{equation}
    f_\gamma\simeq \frac{4\Gamma_a E_\gamma^2}{\pi r \sin\beta m_a^2} \frac{d\dot{N}_a}{dE_a}\left(\frac{m_a^2}{4E_\gamma}\right) \int_1^{+\infty}d\eta \exp\left[-\eta^2\frac{E_\gamma}{T}-\frac{\Gamma_a r \sin \beta}{\eta}\right].
\end{equation}
The dominant contribution to the integral comes from
\begin{equation}
    \eta^{\rm max}= \left(\frac{\Gamma_a r \sin\beta T}{2E_\gamma}\right)^{1/3},
\end{equation}
or from $\eta=1$, depending on whether $\eta^{\rm max}$ is larger or smaller than $1$.
Thus, for $\eta^{\rm max}>1$, the photon distribution function can be approximated as 
\begin{equation}
    f_\gamma\simeq \frac{4\Gamma_a E_\gamma^2}{\pi r \sin\beta m_a^2} \frac{d\dot{N}_a}{dE_a}\left(\frac{m_a^2}{4E_\gamma}\right) \exp\left[-3\left(\frac{\Gamma_a r \sin\beta}{2}\sqrt{\frac{E_\gamma}{T}}\right)^{2/3}\right]\sqrt{\frac{\pi T}{3E_\gamma}},
\end{equation}
while for $\eta^{\rm max}<1$ the correct approximation is
\begin{equation}
    f_\gamma\simeq \frac{4\Gamma_a E_\gamma^2}{\pi r \sin\beta m_a^2} \frac{d\dot{N}_a}{dE_a}\left(\frac{m_a^2}{4E_\gamma}\right) \exp\left[-\frac{E_\gamma}{T}-\Gamma_a r \sin \beta\right]\frac{1}{\frac{2 E_\gamma}{T}-\Gamma_a r \sin\beta}.
\end{equation}

Therefore, for wide angles the angular distribution is non-Gaussian, and the typical scale of suppression is $\sqrt{T/E_\gamma}/\Gamma_a r$. The flux reaching the Earth at the distance $D$ as a function of the angle $\beta$ at large angles is
\begin{equation}
    \frac{d\Phi_\gamma}{dE_\gamma d\beta}\simeq \frac{4\Gamma_a E_\gamma^2 \cos\beta}{\pi r m_a^2} \frac{d\dot{N}_a}{dE_a}\left(\frac{m_a^2}{4E_\gamma}\right) \exp\left[-3\left(\frac{\Gamma_a r \sin\beta}{2}\sqrt{\frac{E_\gamma}{T}}\right)^{2/3}\right]\sqrt{\frac{\pi T}{3E_\gamma}}.
\end{equation}

Finally, we can check the validity of our assumption that the photon emission is stationary. The delay of a photon coming at an angle $\beta$ compared to the photons that reach us directly from the direction of M82 is
\begin{equation}
    \delta t=\frac{\ell}{v_a}+s-r;
\end{equation}
in terms of the angles $\theta$ and $\beta$ we can write it as
\begin{equation}\label{eq:delay}
    \delta t=r(\cos\beta-1)+\frac{r\sin\beta}{\sin\theta}\left(\frac{1}{v_a}-\cos\theta\right),
\end{equation}
since $\ell=r\sin\beta/\sin\theta$ and $s=r\cos\beta-\ell\cos\theta$. If this delay becomes comparable with the typical timescale over which the time distribution of stars in the galaxy changes $\Delta t\sim 1$~Myr (see Fig.~\ref{fig:SFH-IMF}), corresponding to a distance $\Delta t\sim 10^{24}$~cm, then the signal might exhibit a spatial modulation following from the non-stationary emission. The distance from M82 itself is $r\sim 10^{25}$~cm.

From Eq.~\eqref{eq:delay}, we gather that if $\beta$ is very small (our maximum value is $\beta\sim 5'\sim 10^{-3}$) while $\theta\sim 1$ and $v_a\sim 1$, the first term is of order $r\beta^2\ll \Delta t$ while the second term is of order $r\beta\ll \Delta t$. The only possibility to have a sizable delay is if either $v_a\to 0$, which however corresponds to highly non-relativistic axions which only constitute a tiny sliver of the flux and obviously do not affect our results, or if $\theta\sim \beta\ll 1$ (we recall that $\beta$ must be smaller than $\theta$ for the signal to be visible). If $\theta\ll 1$, we can extrapolate its value from Eq.~\eqref{eq:costheta},
\begin{equation}
    \theta\simeq\frac{m_a\sqrt{E_a-E_\gamma}}{E_a\sqrt{E_\gamma}}.
\end{equation}
In order for this to be much smaller than $1$, provided that $E_a$ is not anomalously close to $E_\gamma$ which can only happen for very small regions of the integration space, we must of course have $m_a\ll E_a$, i.e. boosted axions. Therefore, the delay after expanding for small $m_a\ll E_a$ becomes
\begin{equation}
    \delta t\simeq \frac{r\beta^2}{2}+r\frac{m_a \beta}{2\sqrt{E_\gamma(E_a-E_\gamma)}}.
\end{equation}
For very light axions, the second term in the delay is obviously very small, barring again exceptional cases where $E_a$ is very close to $E_\gamma$ which amount to a negligible component of the integration for the total flux. For heavy axions $m_a\gg E_\gamma$, the minimum axion energy as we have seen is $E_a>m_a^2/4E_\gamma$, so the second term is of order $\delta t\sim r\beta\ll \Delta t$. When $m_a$ is comparable with the photon and the axion energies, the second term is still of order $\delta t \sim r\beta\ll \Delta t$. Therefore, we conclude that throughout our range we can always neglect the time dependence in the source emission, albeit the physical reason is different for different mass ranges: for light axions, the large boost impedes the axions to accumulate a sizable delay; for heavy axions, only highly boosted axions with Lorentz factor $\gamma\sim m_a/E_\gamma$ can produce photons in the energy range of NuSTAR, for which again the delay is negligible. Finally, for axions with masses comparable with $100$~keV, the boost factors are not too large compared with unity, but the small angular range considered in this analysis ensures that no sizable delay is accumulated.

\section{C.~Data Reduction and NuSTAR Analysis}
NuSTAR~\cite{NuSTAR:2013yza} is the first space-based focusing  telescope to extend the energy sensitivity into the hard X-ray region, beyond 10-15 keV with observations in the 3-79 keV range. NuSTAR was launched into a near-equatorial low-Earth orbit in 2012.  Each of the two NuSTAR detection lines, focal plane modules A and B (FPMA/FPMB), comprises a telescope module using a segmented-glass approach with 133 nested shells. They are coated with reflective multilayers of different materials (Pt/C for the inner shells, W/Si for the outer ones). The telescope configuration is a conical approximation to a Wolter-I-type geometry designed for grazing-incidence reflection. Each of the two focal plane detectors consist of a 2 by 2 array of CdZnTe (Cadmium Zinc Telluride, also CZT) semiconductor detector chips. Each chip features 32 by 32 pixels, i.e. there are 4096 pixels per detector. With each pixel measuring 0.6 mm $\times$ 0.6 mm, this results in a FOV of about $12'$ in x and y for each focal plane detector at $3$ keV. For higher energies ($ > 10$  keV) the FoV is typically between 6'-10'. The angular resolution of the observatory, primarily determined by its optics, stands at 18" FWHM (Full Width at Half Maximum), with a half-power diameter of about 60''. Designed for optimal performance in the hard X-ray spectrum, NuSTAR's focal plane achieves an energy resolution with a FWHM of 0.4~keV at 10~keV and 0.9~keV at 68~keV.

The dataset examined in this research was retrieved from the NASA's HEASARC\footnote{\href{https://heasarc.gsfc.nasa.gov/cgi-bin/W3Browse/w3browse.pl}{HEASARC Archive web page.}} data archive. The 26 observations considered are detailed in Table \ref{tab:nustar_data} and total an exposure time of $\sim1.96$ Ms.

\begin{table}[h]
\centering
\begin{tabular}{l@{\hspace{6pt}}l @{\hspace{20pt}}|l@{\hspace{6pt}}l @{\hspace{20pt}}|l@{\hspace{6pt}}l}
\toprule
\hline
ObsID          & $t_{\text{exp}}$ [ks] & ObsID          & $t_{\text{exp}}$ [ks] & ObsID          & $t_{\text{exp}}$ [ks] \\
\hline
30101045002    & 189,231              & 30502021002    &  85,157              & 30202022010    &  44,368              \\
50002019004    & 160,740              & 90201037002    &  80,194              & 80202020008    &  40,355              \\
30702012002    & 127,605              & 30502021004    &  78,591              & 30202022002    &  39,021              \\
30901038002    & 123,396              & 30602027002    &  71,868              & 90101005002    &  37,407              \\
30702012004    & 117,506              & 30602027004    &  69,992              & 80202020002    &  36,133              \\
30502022004    &  95,922              & 30602028004    &  68,045              & 80202020004    &  31,667              \\
30502020002    &  88,651              & 30202022004    &  47,035              & 50002019002    &  31,243              \\
30502022002    &  88,348              & 90202038002    &  45,475              & 80202020006    &  30,501              \\
30502020004    &  88,040              & 90202038004    &  43,319              &                &                      \\
\hline
               &                      & Total Exp. [ks]:    & 1959,810              &                &                      \\
\hline
\hline
\end{tabular}
\caption{M82's observation IDs and exposure times used for the NuSTAR data analysis in this work.}
\label{tab:nustar_data}
\end{table}

\begin{figure}[h!]
\centering
\begin{minipage}{.5\textwidth}
  \centering
  \includegraphics[width=\linewidth]{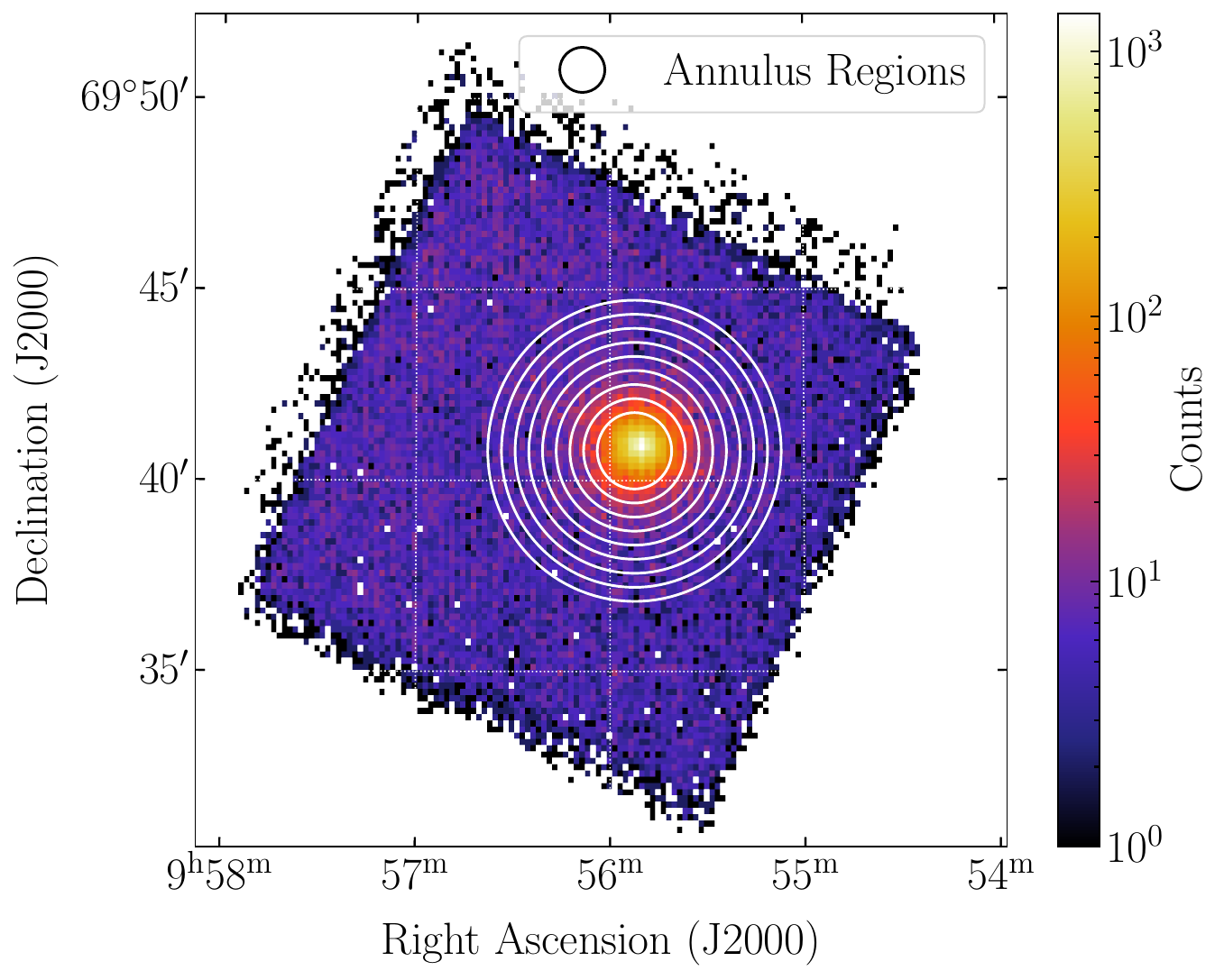}
\end{minipage}%
\begin{minipage}{.5\textwidth}
  \centering
  \includegraphics[width=0.95\linewidth]{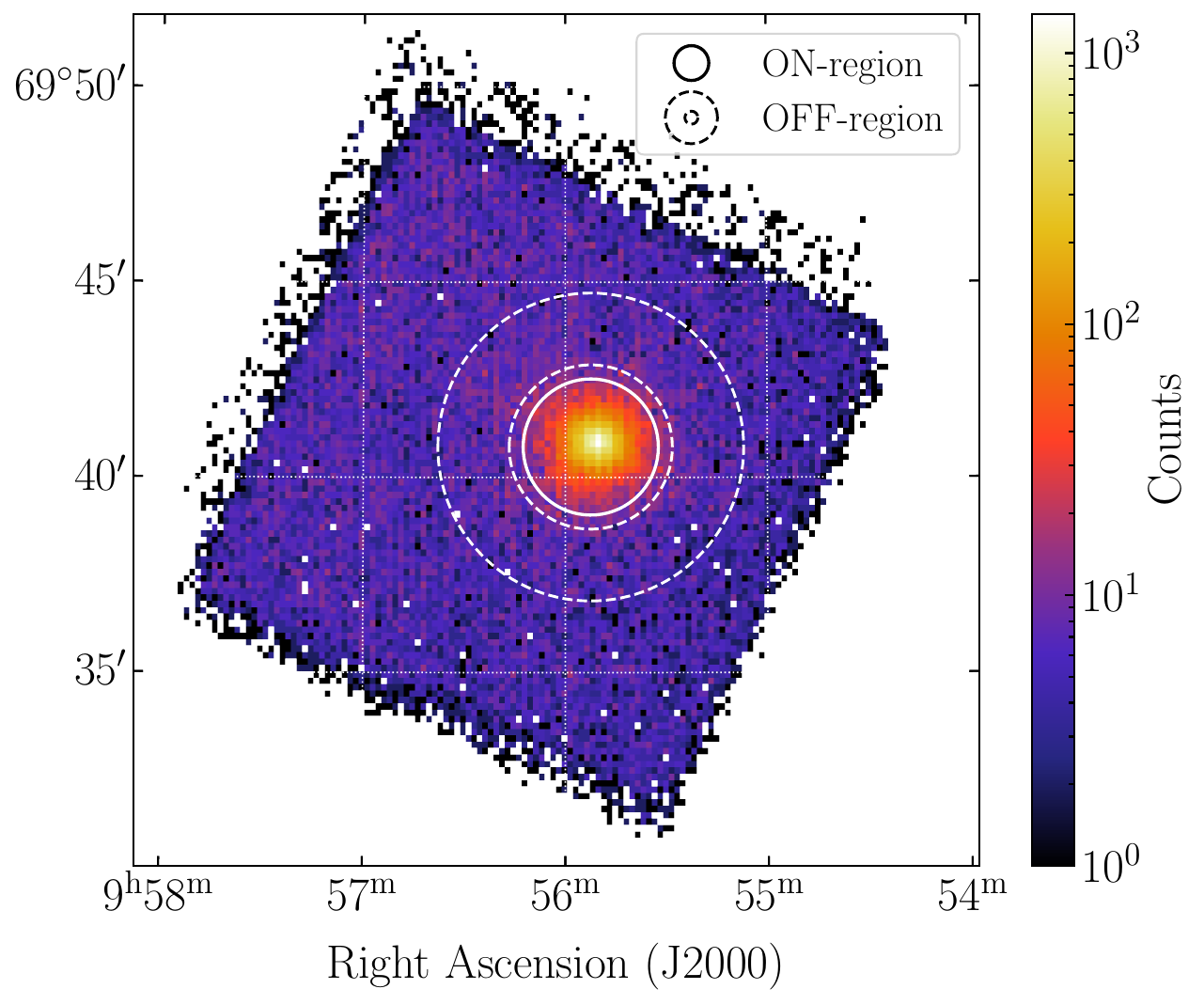}
\end{minipage}
\caption{
Example NuSTAR FOV image (Obs. ID 30202022010A). Left: Concentric annuli around the bright spot delineate the regions used in our main analysis, where an isotropic bakground was assumed. Right: The ON and OFF regions used for the background subtraction analysis in the SupM are highlighted.}
\label{fig:nu_star_fov}
\end{figure}

To process the NuSTAR observations, we employed the NuSTARDAS software suite (version 2.1.4), which is part of HEASoft package (version 6.34) \cite{ascl:1408.004}. The workflow begins with reprocessing the raw data from the archival files using the \texttt{nupipeline} task provided by NuSTARDAS. As explained in the main text and shown in the left panel of Figure \ref{fig:nu_star_fov}, the source regions are defined as follows: an inner circular region with a radius of 60'' centered on the coordinates of M82, and eight concentric annular regions extending from 60'' to 237''. Each annulus is separated by a uniform radial distance of 22'', consistent with the telescope's FWHM of 18''.
The 60'' source region captures most of the M82 X-ray emission. To ensure that all regions remain within the detector, the maximum radius for the outermost annulus is set to 237'', accounting for the fact that the source is not perfectly centered within the FOV in all observations. 

\begin{figure}[h!]
    \centering
    \includegraphics[width=1\linewidth]{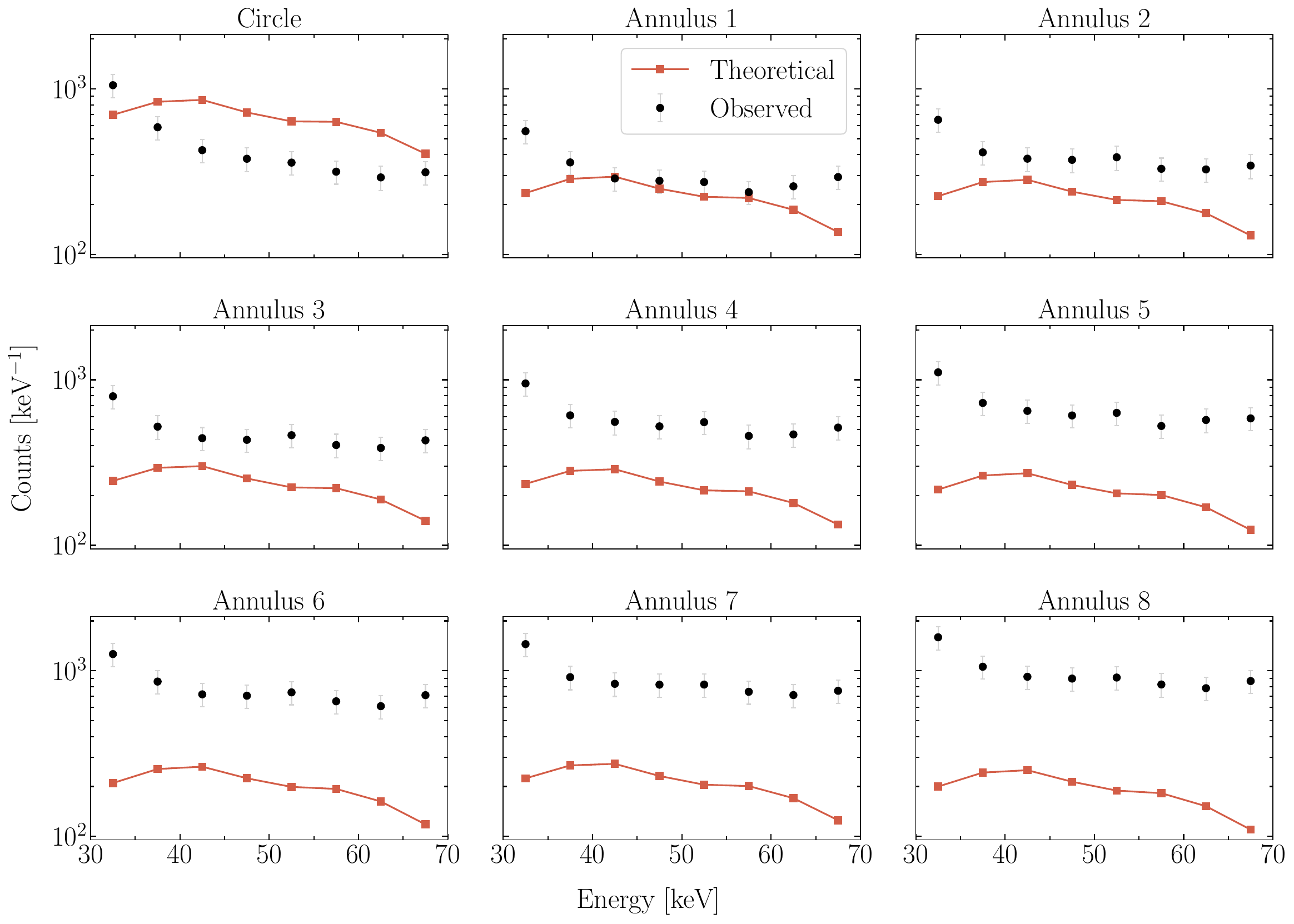}
    \caption{
The stacked and binned spectra, which combine contributions from FPMA and FPMB and all observations, are shown for the observational data (black dots) alongside the predicted axion-induced spectra (red curve) corresponding to an example parameter set: $\boldsymbol{\theta} = \{g_{a\gamma} = 10^{-12}\ \mathrm{GeV}^{-1},\ m_a = 100\ \mathrm{keV}\}$.}
    \label{fig:angular_spectra}
\end{figure}

Next, we use \texttt{nuproducts} with the previously mentioned regions across all observations for both modules A and B, generating a total of 468 spectra (2 FPMs, 9 regions, 26 observations). After obtaining the spectra, we utilize the \texttt{grppha} tool from \texttt{FTOOLS} \cite{blackburn1995ftools} version 6.34 to rebin the data into 5 keV bins within the range of 30 keV to 70 keV.
Then, we employ HEASoft's \texttt{XSPEC} tool \cite{arnaud1999xspec} version 12.14.1 to read the spectra and export the data into ASCII format. Finally, we aggregate all spectra for each annulus, producing a total of nine combined spectra. These aggregated spectra are depicted as black dots in Figure \ref{fig:angular_spectra}.

On the other hand, the predicted differential photon flux from axion decay is forward-modeled with the response files for each of the two optics and detector (Ancillary Response Files (ARFs) and Response Matrix Files (RMFs), respectively) to determine the expected counts at the NuSTAR detector. Thus, the axion-induced expected counts for an observation $\varepsilon$, energy bin $i$, and annulus $\alpha$ are given by
\begin{equation}
   N^a_{\varepsilon, i, \alpha}(g_{a\gamma}, m_a) = t^\varepsilon \int dE' \, \mathrm{RMF}_{i, \alpha}^\varepsilon(E_i,E') \, \mathrm{ARF}^\varepsilon_{\alpha}(E') \, S_{\alpha}(E' \mid g_{a\gamma}, m_a).
\end{equation}
where $t^{\varepsilon}$ corresponds to the exposure time of the observation $\varepsilon$ and $\{g_{a\gamma}, m_a\}$ denotes the parameters to fit, axion-photon coupling and axion mass. Subsequently, the expected counts are combined as $N^a_{i, \alpha} = \sum_{\varepsilon} N_{\varepsilon, i, \alpha}$, resulting in nine distinct spectra.These spectra, represented by red squares in Figure \ref{fig:angular_spectra}, correspond to each annulus and will be incorporated into our likelihood analysis.

\section{D.~Background treatment}

\begin{figure*}
    \includegraphics[width=\textwidth]{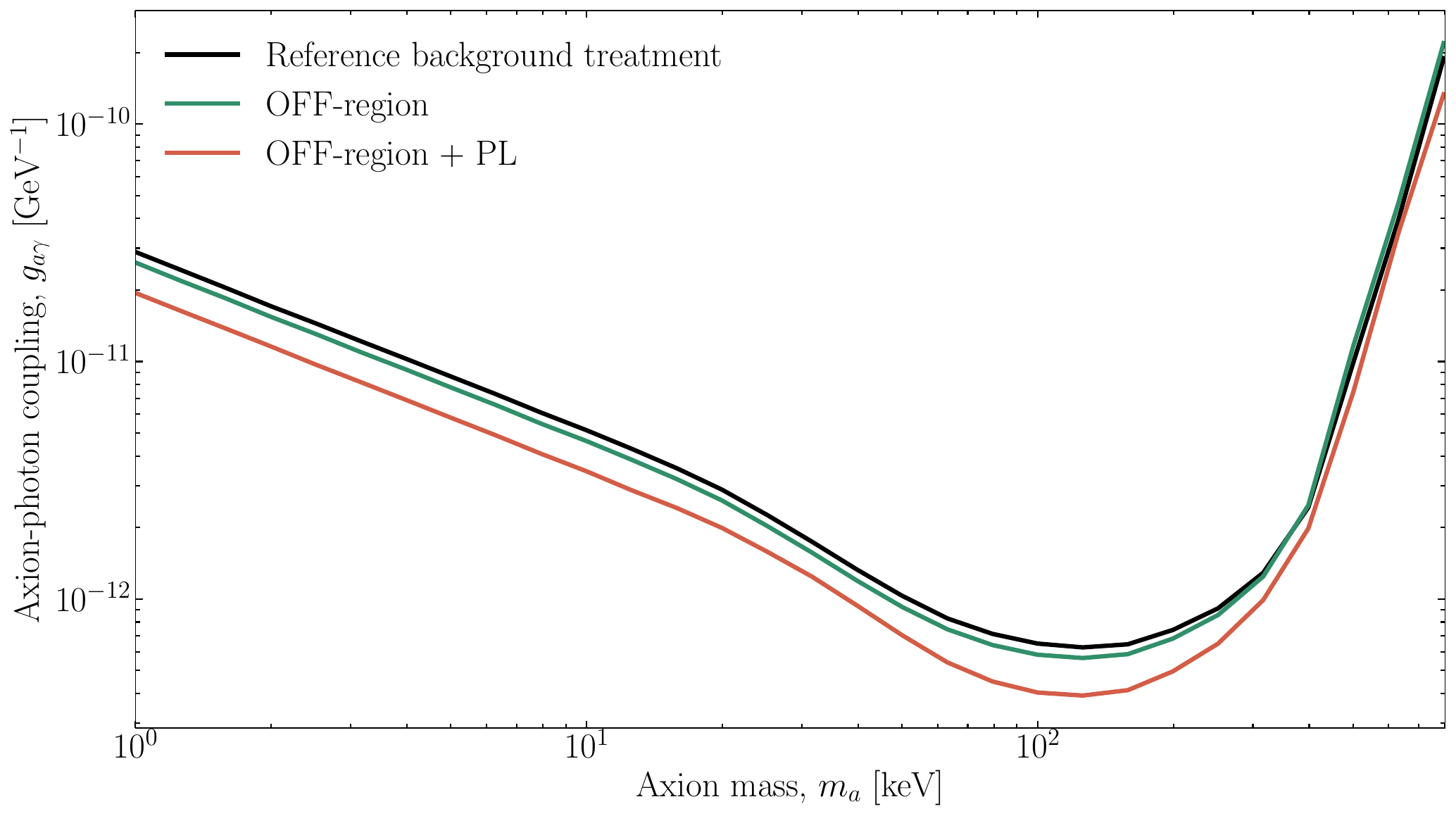}
    \caption{Constraints derived from three different background treatments are shown: (i) the main analysis using a simplified isotropic background, (ii) the OFF-region background subtraction method, and (iii) the OFF-region method with an added power-law component to account for M82’s intrinsic X-ray emission. The results indicate that subtracting the OFF-region and including the power-law model further tighten the constraints, affirming the conservative nature of our primary analysis.}\label{fig:bg}
\end{figure*}

In the main text, we have obtained our constraints based on a very conservative approach, assuming as the only components an isotropic background and the axion signal. Note that this approach does not fully account for the detector’s realistic background, which in practice includes additional non-isotropic components (see, e.g. \cite{Wik_2014,Krivonos2021}) as well as the intrinsic X-ray emission from the M82 galaxy. Our restriction to this simplified approach was primarily motivated by its conservative nature: by neglecting additional background components, we ensure that the axion signal is as weakly constrained as possible. This also allows us to avoid an ad-hoc choice of an OFF-region to identify the background, which is convenient since the axion signal is spread over a range that depends on the coupling we are constraining. In this Section we verify the conservativeness of this approach, by testing different strategies to model the background, and verifying that the constraints we obtained by these different strategies are stronger.

Rather than self-consistently modeling the background, we can proceed with identifying an OFF-region, to provide us with an estimate of the background, and an ON-region centered around M82. Motivated by the typical widths of the axion signal for the values of coupling at which we find our constraints, we choose as the ON-region the angular region composed of the three central bins used in the text, i.e. a circle of angular opening $1.74'$, whereas as OFF-region we use the region which includes the angular bins from $5$ to $9$, i.e. an annulus between the opening angles $2.1'$ and $3.94'$ (see the right panel of Figure \ref{fig:nu_star_fov}). The background is extracted from the OFF-region using the standard NuSTAR analysis procedure and is automatically rescaled to the source area by \texttt{nuproducts}. Therefore, we obtain the constraints by fitting the observed counts from the ON-region with either \textit{i)} the fixed background estimated from the OFF-region and the axion signal, or \textit{ii)} the fixed background, the axion signal, and the photons from the central M82 source, modeled as a power law (PL) with free normalization and spectral index. The latter approach corresponds to the one followed in Ref.~\cite{Ning:2024eky}.

Figure~\ref{fig:bg} shows the resulting constraints. The ``Reference background treatment'' (black line) corresponds to our main text approach, whereas the approaches ``OFF-region'' (green) and ``OFF-region + PL'' (red) correspond to \textit{i)} and \textit{ii)} respectively. As expected, when the background is estimated from the OFF-region, rather than self-consistently fit under the assumption of isotropy, the constraints are very slightly more aggressive, because of the reduced flexibility with which the axion and background signal can be fit to the observations. A slightly more significant improvement is found once a power-law component is introduced in the fit, so that the axions cannot overshoot the fluctuations of the data on top of the expected power-law signal. Such an improvement becomes smaller for heavier axions ($m_a\gg 100$~keV). Indeed, for these masses constraints are dominated by the highest-energy bins considered in our analysis, where the PL contribution is suppressed and the astrophysical background from the central M82 source becomes smaller. These different lines can be interpreted as the characteristic uncertainty due to the background modeling on our constraints. The bounds we show in the main text are the most conservative among these constraints.

\section{E.~Constraints compared to the general parameter space}

\begin{figure}
    \centering
    \includegraphics[width=\textwidth]{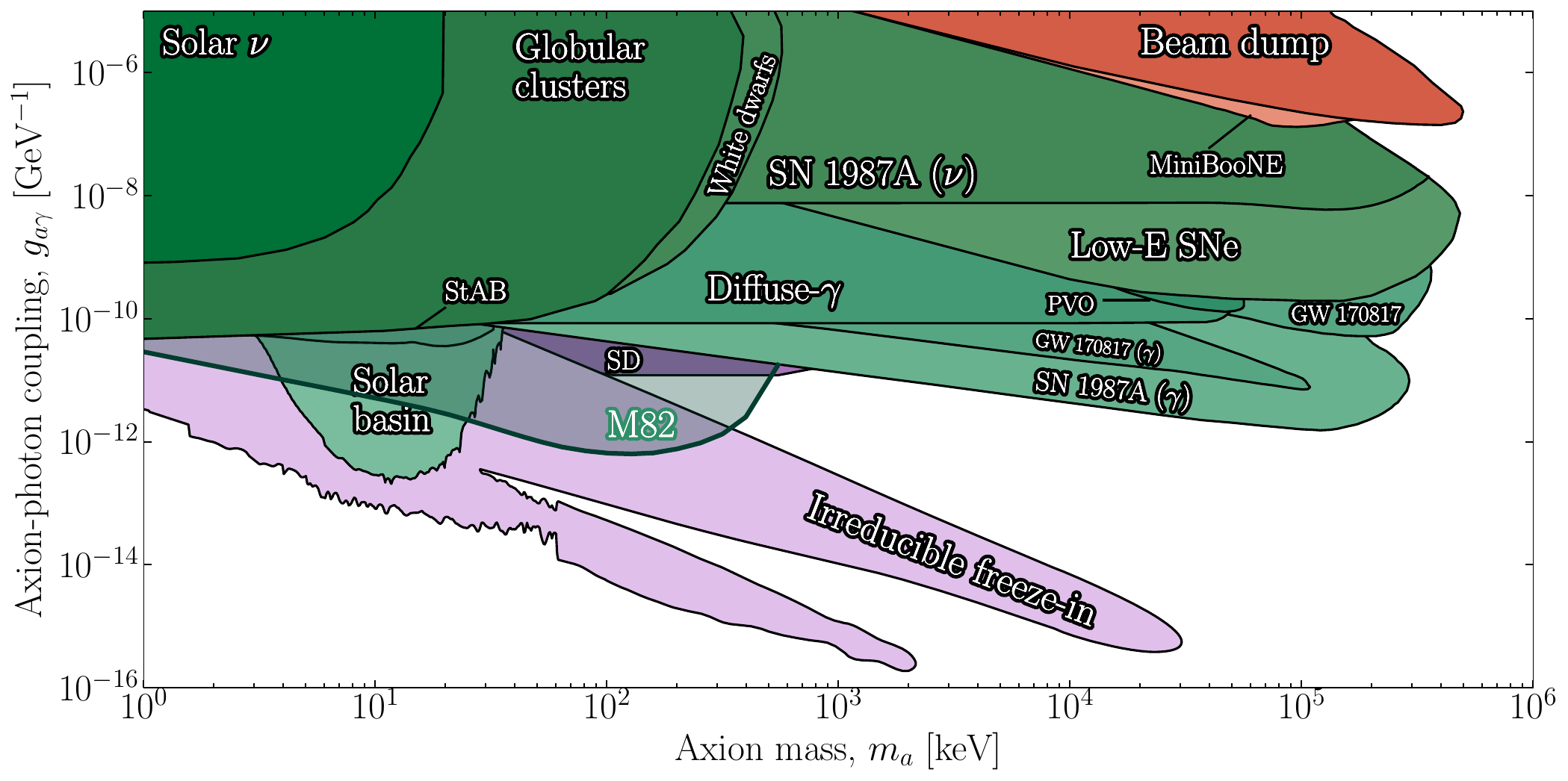}
    \caption{New constraints on axion decay from M82 in the parameter space of axion mass $m_a$ and axion-photon coupling $g_{a\gamma}$, compared with the pre-existing constraints. We show Earth-based bounds in red~\cite{CHARM:1985anb,Riordan:1987aw,Blumlein:1990ay,NA64:2020qwq,Dolan:2017osp,Capozzi:2023ffu}, astrophysical bounds in green~\cite{Vinyoles:2015aba,Dolan:2021rya,Caputo:2021rux,Caputo:2022mah,Dolan:2022kul,
    Hoof:2022xbe, Nguyen:2023czp,Diamond:2023cto,Diamond:2023scc,Dev:2023hax,Beaufort:2023zuj} (see Ref.~\cite{Fiorillo:2023frv} for a recent reassessment of SN~1987A neutrino cooling), and cosmological bounds in purple~\cite{Balazs:2022tjl,Langhoff:2022bij}; a collection of these bounds can be found at Ref.~\cite{AxionLimits}. The new bounds from M82 are shown in dark green.}
    \label{fig:fig1_zoom}
\end{figure}

In the main text, we have compared our new constraints with the pre-existing astrophysical and cosmological bounds in the mass range between $10$~keV and $1$~MeV. Figure~\ref{fig:fig1_zoom} shows the general constraints in the parameter space for the axion-photon coupling across the entire mass range from $1$~keV to $1$~GeV. At low masses, below $m_a\lesssim 10$~keV, the bounds approximately follow $g_{a\gamma}\propto m_a^{-1}$. This behavior can be qualitatively understood noting that at low masses the axion spectrum is essentially mass-independent, but for sufficiently low couplings the fraction of the photon spectrum within the angular region considered is proportional to $f_\gamma\propto \frac{d\dot{N}_a}{dE_a}\Gamma_a m_a$ (see Eq.~\eqref{eq:photon_distribution} in the limit $\Gamma_a m_a r/p_a \sin\theta\ll 1$), and since $\Gamma_a\propto g_{a\gamma}^2 m_a^3$ we see that the photon signal grows with $g_{a\gamma}^4 m_a^4$. On the other hand, since the angular region dominating the constraints is close to the center, but not precisely at the center, where the signal-to-noise ratio is highest, the precise dependence of $g_{a\gamma}$ on $m_a$ is difficult to guess with simple arguments. At large masses $m_a\gg 300$~keV, the constraints weaken due to the kinematic suppression of axion production in the stars. Overall, our new constraints are by far the dominant ones compared to the previous astrophysical bounds, essentially mirroring the supernova bounds at higher masses, and they exclude new parameter space even compared with the cosmological bounds by as much as an order of magnitude.

\end{document}